\documentclass[lettersize,journal]{IEEEtran}
\usepackage{amsmath,amsfonts}
\usepackage{algorithmic}
\usepackage{algorithm}
\usepackage{array}
\usepackage{float}
\usepackage[caption=false,font=normalsize,labelfont=sf,textfont=sf]{subfig}
\usepackage{textcomp}

\usepackage{stfloats}
\usepackage{url}
\usepackage{tikz}
\usetikzlibrary{trees}
\usepackage{verbatim}
\usepackage{graphicx}
\usepackage{cite}
\usepackage{tabularx}
\usepackage{booktabs}
\usepackage{array}
\usepackage{braket}
\usepackage{multicol}
\usepackage{multirow}
\usepackage{makecell}

\begin{document}

\title{Towards Adapting Federated \& Quantum Machine Learning for Network Intrusion Detection: A Survey}

\author{Devashish Chaudhary, Sutharshan Rajasegarar and Shiva Raj Pokhrel
\thanks{The authors are with the School of Information Technology, Deakin University, Geelong, Australia (e-mail: s224281473@deakin.edu.au; sutharshan.rajasegarar@deakin.edu.au; shiva.pokhrel@deakin.edu.au)}}



\maketitle

\begin{abstract}

This survey explores the integration of Federated Learning (FL) with Network Intrusion Detection Systems (NIDS), with particular emphasis on deep learning and quantum machine learning approaches. FL enables collaborative model training across distributed devices while preserving data privacy—a critical requirement in network security contexts where sensitive traffic data cannot be centralized. Our comprehensive analysis systematically examines the full spectrum of FL architectures, deployment strategies, communication protocols, and aggregation methods specifically tailored for intrusion detection. We provide an in-depth investigation of privacy-preserving techniques, model compression approaches, and attack-specific federated solutions for threats including DDoS, MITM, and botnet attacks. The survey further delivers a pioneering exploration of Quantum FL (QFL), discussing quantum feature encoding, quantum machine learning algorithms, and quantum-specific aggregation methods that promise exponential speedups for complex pattern recognition in network traffic. Through rigorous comparative analysis of classical and quantum approaches, identification of research gaps, and evaluation of real-world deployments, we outline a concrete roadmap for industrial adoption and future research directions. This work serves as an authoritative reference for researchers and practitioners seeking to enhance privacy, efficiency, and robustness of federated intrusion detection systems in increasingly complex network environments, while preparing for the quantum-enhanced cybersecurity landscape of tomorrow.

\end{abstract}


\section{Introduction}

\IEEEPARstart{W}{ith} the rapid evolution in modern networks like IoT devices, 5G technologies, and cloud services \cite{10112761,dangi2021study,rao2021evolving}, it has been important to maintain the security of these networks. These networks facilitate huge amounts of sensitive and confidential data sharing, making their security critical. Any compromise to the data can lead to significant consequences. For instance, the lawsuit against Ring highlights the consequences of data leaks, where inadequate security measures allowed hackers to exploit smart home devices, leading to harassment, threats, and severe privacy violations \cite{theguardian2020amazon}. Network Intrusion Detection System (NIDS) plays a crucial role in safeguarding the network data and identifying any malicious activities within the network \cite{ahmad2021network}. Traditional IDS are trained using a centralized approach, where all the data are transferred to a centralized system for model training. This encounters significant limitations due to large bandwidth requirements for communication, experiencing detection delays and privacy consequences \cite{khraisat2019survey}. This limits the IDS's capability to timely detect evolving cyber risks and highly refined intrusions.

Communication plays an important role in a network of devices, particularly in IoT networks. The interconnectedness of IoT devices is expected to reach 40 billion by 2030 \cite{iot_devices}. Identifying anomalous behavior in wireless sensor networks must balance accuracy with energy and communication efficiency~\cite{rajasegarar2008anomaly}. Prior distributed approaches have attempted to address this challenge by reducing communication overhead~\cite{rajasegarar2006distributed,rajasegarar2014ellipsoidal}, yet they still required sharing processed data across nodes. Federated Learning (FL) advances this idea by eliminating the need to transfer even processed sensor measurements. Instead, only model parameters are communicated with the server, thereby further reducing communication costs while preserving data privacy compared to traditional centralized methods~\cite{zhou2021communication, dritsas2025federated}. Though FL techniques require the frequent transmission of local model updates, various techniques such as model compression \cite{le2024survey}, asynchronous updates \cite{wang2022asynchronous} and communication efficient aggregation \cite{bonawitz2019federated} can be used to mitigate communication bottlenecks. In the centralized method, the model is trained only when all the actual raw data from end devices are transferred to the server. However, IoT devices often operate in remote geographical locations with limited computational power, energy and network connectivity, which can lead to communication failures, preventing some devices from sending their data to the server for training~\cite{o2014anomaly}.

\begin{figure}[t]
    \centering
    \includegraphics[width=0.80\linewidth]{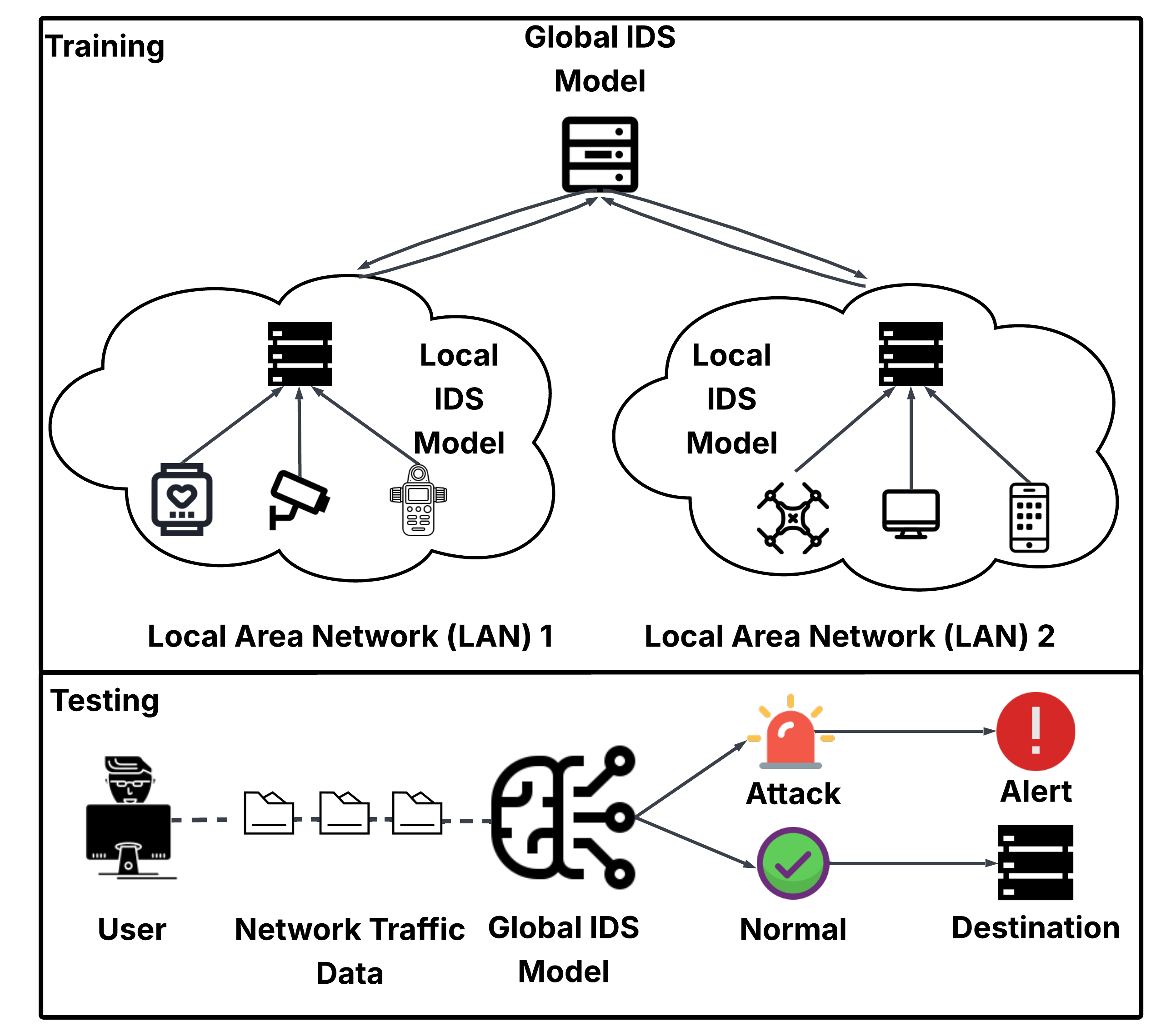}
    \caption{Overview of Federated Learning for Intrusion Detection: Illustration of the training phase, where local models are trained on distributed datasets at multiple devices without sharing raw data, and the testing phase, where the aggregated global model is evaluated for detecting network intrusions.}
    \label{fig:fl_overview}
\end{figure}
To address these challenges, the FL approach has been introduced \cite{li2020review}. FL can be used to produce a global model that is dynamically updated. Fig.~\ref{fig:fl_overview} shows a high level overview of FL in IDS. When an end device experiences a loss of network or energy, it can defer transmitting its local model updates. Once the device regains connectivity and sufficient power, it can send its updates which are then dynamically incorporated into the global model \cite{mughal2024adaptive}. The global model \cite{qi2024model} produced by FL synthesizes knowledge from all the participating clients, capturing insights and patterns from their diverse data while maintaining data privacy. 

There are various challenges in analyzing the data that are generated across multiple distributed nodes (data sources) \cite{zhang2021survey}, including:
\begin{itemize}
    \item Scalability Issues: Processing a high volume of data from diverse geographical locations can lead to bottlenecks in performance.
    \item Non-IID Data: Data generated by different edge nodes exhibit non-identical and independent distributions (non-IID), which makes the training of a unified model challenging.
    \item Fault Tolerance: The nodes may disconnect or provide incomplete data which can make the collected data unreliable.
    \item Computational Constraints on Edge Devices: The nodes are usually devices that have less computational power and operate on batteries. Analyzing huge amounts of data on the nodes itself becomes computationally challenging.
    \item Dynamic Network Topologies: Real-time network intrusion detection has always been a challenge due to ever-evolving traffic patterns, network configurations, and device connections.
\end{itemize}

Intrusion detection often involves analyzing sensitive data, such as user activities, communication patterns, and system logs. Centralized approaches require these sensitive raw data to be sent to the central server for processing, which poses significant privacy risks. These challenges necessitate novel approaches, such as FL, which decentralize the analysis process, maintain data privacy, and address the unique constraints of distributed networks.

Traditional machine learning models, such as support vector machine (SVM), K-nearest neighbor (KNN), naïve Bayes, logistic regression (LR), decision trees, clustering, and combined and hybrid methods have been widely used in NIDS. However, machine learning (ML) models often rely on handcrafted features, requiring domain expertise and extensive pre-processing to ensure optimal performance \cite{liu2019machine}. This manual feature engineering can limit the scalability and adaptability of traditional ML approaches in dynamic and complex network environments. Deep Learning (DL), on the other hand, has shown great accuracy in detecting sophisticated network attacks. It automates feature extraction and leverages hierarchical representations of data. DL models can generalize better in heterogeneous network conditions and dynamically adapt to evolving attack patterns \cite{liu2019machine}. This makes them a powerful alternative to traditional ML approaches for NIDS, especially in environments where accuracy, scalability, and minimal human intervention are critical. While DL often requires more computational resources than ML, advancements in hardware and distributed frameworks like FL have mitigated these challenges, further enhancing the viability of deep learning for modern NIDS.

The research in QML is gaining popularity as it can be used to handle high-dimensional complex network traffic data~\cite{amin2018quantum}. Although DL methods can detect intricate patterns, they often struggle as the volume and dimensionality of the data increase. QML, leveraging quantum phenomena such as superposition and entanglement, offers a new computational paradigm that could provide significant advantages in representational power and learning efficiency. In particular, researchers are exploring hybrid quantum-classical methods that could provide significant advantages in representing high-dimensional complex network data and increasing learning efficiency. However, in the real world, the practical deployment of QML is currently constrained due to limited access to actual quantum hardware. Despite this, early research indicates that QML may hold the key to building more efficient and robust IDS in the future.

While existing surveys on federated learning have highlighted its applications, this survey article provides a comprehensive and up-to-date study of the application of \textit{FL with deep learning} and \textit{QML}, particularly focused on Intrusion Detection Systems (IDS). The key contributions are as follows:
\begin{enumerate}
    \item Provides the first comprehensive analysis integrating classical federated learning, deep learning, and quantum approaches for network intrusion detection
    \item Presents a systematic taxonomy of federated learning architectures, deployment strategies, and aggregation methods specifically tailored for network security
    \item Offers an in-depth exploration of quantum federated learning techniques that promise exponential speedups for complex pattern recognition in network traffic
    \item Identifies critical research gaps and outlines a roadmap for industrial adoption of federated intrusion detection systems
    \item Proposes future research directions addressing privacy, efficiency, and robustness challenges in increasingly complex network environments
\end{enumerate}
We highlight key concepts, system designs, strategies, attack-specific existing solutions, challenges, and future possibilities, giving a well-rounded view of how FL can enhance IDS using DL and QML. The article discusses the basics of FL, practical approaches for architecture and communication in IDS, and methods for ensuring privacy, security, and resilience. It also highlights the latest research trends and innovations in the field. To assist readers in navigating the technical terminology used throughout this survey, Table~\ref{tab:glossary} provides a glossary of the key acronyms and abbreviations in alphabetical order.

\begin{table}[ht]
\centering
\caption{Glossary of key acronyms and abbreviations used in this survey.}
\begin{tabular}{l p{0.7\linewidth}} 
\hline
\textbf{Acronym} & \textbf{Full Form} \\
\hline
API & Application Programming Interface \\
AUC & Area Under the Curve \\
BQP & Bounded-error Quantum Polynomial time \\
CNN & Convolutional Neural Network \\
DDoS & Distributed Denial of Service \\
DL & Deep Learning \\
DNN & Deep Neural Network \\
DP & Differential Privacy \\
FATE & Federated AI Technology Enabler \\
FedAvg & Federated Averaging \\
FedProx & Federated Proximal \\
FedSVRG & Federated Stochastic Variance Reduced Gradient \\
FL & Federated Learning \\
FL-DAD & Federated Learning for Decentralized DDoS Attack Detection \\
FL-LSTM & Federated Learning with Long Short-Term Memory \\
GNN & Graph Neural Network \\
GRU & Gated Recurrent Unit \\
HFL & Horizontal Federated Learning \\
HIDS & Host-based Intrusion Detection System \\
IDS & Intrusion Detection System \\
IoT & Internet of Things \\
KNN & k-Nearest Neighbour \\
LR & Logistic Regression \\
ML & Machine Learning \\
MITM & Man-in-the-Middle \\
NIDS & Network Intrusion Detection System \\
NP & Nondeterministic Polynomial time \\
P & Polynomial time \\
PCA & Principal Component Analysis \\
QC & Quantum Computing \\
QAE & Quantum Autoencoder \\
QBM & Quantum Boltzmann Machine \\
QDP & Quantum Differential Privacy \\
QEC & Quantum Error Correction \\
QFL & Quantum Federated Learning \\
QGAN & Quantum Generative Adversarial Network \\
QHE & Quantum Homomorphic Encryption \\
QKD & Quantum Key Distribution \\
QkNN & Quantum k-Nearest Neighbour \\
QML & Quantum Machine Learning \\
QNN & Quantum Neural Network \\
QOTP & Quantum One-Time Pad \\
QPCA & Quantum Principal Component Analysis \\
QPU & Quantum Processing Unit \\
QRL & Quantum Reinforcement Learning \\
QSVC & Quantum Support Vector Classifier \\
QSVM & Quantum Support Vector Machine \\
RNN & Recurrent Neural Network \\
RL & Reinforcement Learning \\
SCAFFOLD & Stochastic Controlled Averaging for Federated Learning \\
SDN & Software-Defined Network \\
SHAP & SHapley Additive exPlanations \\
SMPC & Secure Multi-Party Computation \\
SVM & Support Vector Machine \\
TFF & TensorFlow Federated \\
XAI & Explainable Artificial Intelligence \\
\hline
\end{tabular}
\label{tab:glossary}
\end{table}

The remainder of this paper is organized as follows. Section II introduces the necessary background and fundamentals, including intrusion detection systems, deep learning in IDS, and federated learning. Section III surveys related works, while Section IV provides a comprehensive review of classical federated learning approaches, covering types of FL, communication methods, deployment strategies, model aggregation, privacy preservation, network protocols, available frameworks, model compression techniques, and attack-specific solutions. Section V shifts focus to QML, discussing quantum gates and circuits, feature encoding, applications of QML in IDS, and representative case studies. Section VI extends this discussion to quantum federated learning (QFL), examining deployment, communication, secure parameter sharing, and quantum-specific aggregation methods. Section VII discusses standardized evaluation metrics in intrusion detection. Section VIII outlines key challenges. Section IX addresses regulatory, ethical, and societal considerations. Section X and Section XI highlight the roadmap for industrial adoption of IDS and promising future directions, respectively. Finally, Section XII concludes the paper.

\section{Background and Fundamentals}
\subsection{Intrusion Detection System (IDS)}
    In the realm of network security, ensuring the confidentiality, integrity, and availability of data is important. IDS are critical in safeguarding networks and systems against malicious activities and unauthorized access \cite{liao2013intrusion}. An IDS is a security solution that monitors and analyzes network traffic, system logs, and activities to detect potential threats, such as unauthorized access, malware, and policy violations.

    IDS can be broadly classified, based on their monitoring strategies and deployment, into the following categories \cite{khraisat2019survey,liao2013intrusion}:
            \begin{itemize}
                \item Host-Based IDS (HIDS): HIDS operates at the host level, monitoring activities such as file integrity, process execution, and system logs. These systems are particularly effective in detecting insider threats and anomalies specific to a single system.
        
                \item Network-Based IDS (NIDS): NIDS monitors network traffic and analyzes packet data to identify suspicious patterns or anomalies. These systems are deployed at strategic points within the network, such as gateways or routers, to ensure comprehensive traffic monitoring.
        
                \item Hybrid IDS: Hybrid systems combine the capabilities of HIDS and NIDS, providing an integrated approach to intrusion detection by monitoring both network traffic and host-based activities.
            \end{itemize}

    NIDS are classified into two categories based on their detection methodology: signature-based detection and anomaly-based detection. Traditional IDS are based on signature-based anomaly detection and they use the traditional machine learning algorithms to detect anomalies in network traffic. Signature-based detection relies on predefined patterns, known as signatures, of known attacks. These signatures are essentially fingerprints of malicious activities, derived from historical data and prior analysis \cite{alazab2021federated}. The system matches incoming network traffic against this signatures database to identify intrusions. Though this traditional method can detect known attacks/anomalies with higher accuracy, it is inefficient in detecting unknown anomalies or zero-day attacks \cite{liao2013intrusion}. As network data continues to evolve with emerging threats and novel attack patterns, there is a critical need for an alternative approach to address these challenges. This is where anomaly-based detection becomes essential, providing a method to identify deviations from normal behavior that may indicate potential attacks \cite{alazab2021federated}. This method leverages statistical models, machine learning, or deep learning to characterize normal traffic and flag any deviations as potential intrusions. It is very efficient in detecting zero-day attacks. The main disadvantage of this method is that it can lead to high false positives \cite{liao2013intrusion}. It also requires large volumes of normal behavior data for training, which can be challenging to obtain. Combining signature and anomaly-based techniques, hybrid systems aim to achieve a balance between accuracy and adaptability, improving detection rates while minimizing false alarms.

\subsection{Deep Learning in NIDS}
NIDS are essential tools in cyber security, that are used to identify and respond to malicious activities within a network \cite{ahmad2021network}. Traditional NIDS techniques, such as signature-based and rule-based methods, have limitations in detecting unknown or evolving attack patterns. To address these limitations, deep learning (DL), a powerful subset of machine learning, has gained significant attention for enhancing NIDS performance.
Deep learning models like Convolutional Neural Networks(CNNs), Recurrent Neural Networks(RNNs) and Autoencoders have shown promising results in NIDS \cite{chen2018autoencoder,mohammadpour2022survey,tang2018deep}. CNNs are effective at identifying spatial patterns in network traffic \cite{kim2020cnn}. RNNs are developed to process sequential data to capture temporal attack behaviors \cite{al2007power}, and Autoencoders are particularly useful for anomaly detection by learning the normal patterns of network activity \cite{chen2018autoencoder}.

The key advantages of deep learning for NIDS include the ability to automatically extract features from raw traffic data, handle large-scale datasets, and detect previously unknown attack vectors. Despite these benefits, challenges remain, such as the need for large, labeled datasets, high computational requirements, and addressing class imbalance in network traffic.

Recent advancements in deep learning for NIDS have explored hybrid models, combining the strengths of traditional machine learning with deep learning techniques \cite{sajid2024enhancing}, as well as innovations like reinforcement learning, model compression techniques and federated learning \cite{agrawal2022federated}, which offer solutions to data privacy and resource constraints.

\subsection{Foundations of Federated Learning}
This section presents an overview of FL and how it can be implemented efficiently in NIDS. This section also analyzed the current survey papers on FL, particularly focusing on recent trends and future directions for NIDS.

\subsubsection{Concept}

FL is based on client-server architecture, where several clients train a local model on their local data (see Fig.~\ref{fig:Federated_Learning}(a)). Unlike the traditional centralized approach, where the raw data needs to be transferred to the central server from all the clients for model training, only model parameters are shared with the central server for aggregation. The central server is responsible for aggregating the parameters from all the clients and producing a global model which is again sent back to the clients for further training on any newer data available. This process helps in collaborative training among the clients and also helps to produce a robust model that will be able to detect sophisticated attacks, both at the local and global levels.

In centralized learning, all data from the clients are transferred to a central server for model training. Let the dataset of the $n$ clients be represented as \( D_1 \cup D_2 \cup \dots \cup D_n \). The central server trains a global model \( \theta_{\text{centralized}} \) on the combined dataset: 
\[ \theta_{\text{centralized}} = \arg \min_{\theta} \mathcal{L}(\theta, D_1 \cup D_2 \cup \dots \cup D_n), \] where \(\mathcal{L}(\theta, D)\) is the loss function over the combined dataset, and \(\theta\) represents the model parameters.

In federated learning, each client keeps its data locally and trains its own model. Let the dataset of the \( i \)-th client be denoted by \( D_i \), and each client trains a local model \( \theta_i \) on its local dataset: 
\[\theta_i = \arg \min_{\theta} \mathcal{L}(\theta, D_i). \]

Once the local models are trained, they are sent to the central server for aggregation. The global model \( \theta_{\text{global}} \) is updated by aggregating the local models \( \theta_1, \theta_2, \dots, \theta_n \), but the exact method of aggregation is deferred to another section. The central server then sends the updated global model back to the clients for further local training: 
\[ \theta_{\text{global}} = \text{Aggregate}(\theta_1, \theta_2, \dots, \theta_n).
\]

The main objective of federated learning is to achieve performance that is as close as possible to centralized learning while addressing the challenges of data privacy and security \cite{yang2019federated}. In mathematical terms, this means minimizing the difference between the loss function of the global model in both approaches. The objective can be expressed as:
\[
 \min_{\theta_{\text{global}}} \left| \mathcal{L}(\theta_{\text{global}}, D_{\text{local}}) - \mathcal{L}(\theta_{\text{centralized}}, D_{\text{combined}}) \right|,
\]
here, \( \theta_{\text{global}} \) refers to the global model obtained from federated learning, and \( \theta_{\text{centralized}} \) refers to the global model in centralized learning. \( D_{\text{local}} \) represents the local datasets in federated learning, and \( D_{\text{combined}} \) is the combined dataset in centralized learning. The goal is to minimize the difference between the losses of the models in both settings, ensuring that federated learning performs as effectively as centralized learning, despite its decentralized nature.

\subsubsection{Federated Learning Life cycle}

\begin{figure*}[ht]
    \centering
    \subfloat[]{\includegraphics[width=0.349\linewidth]{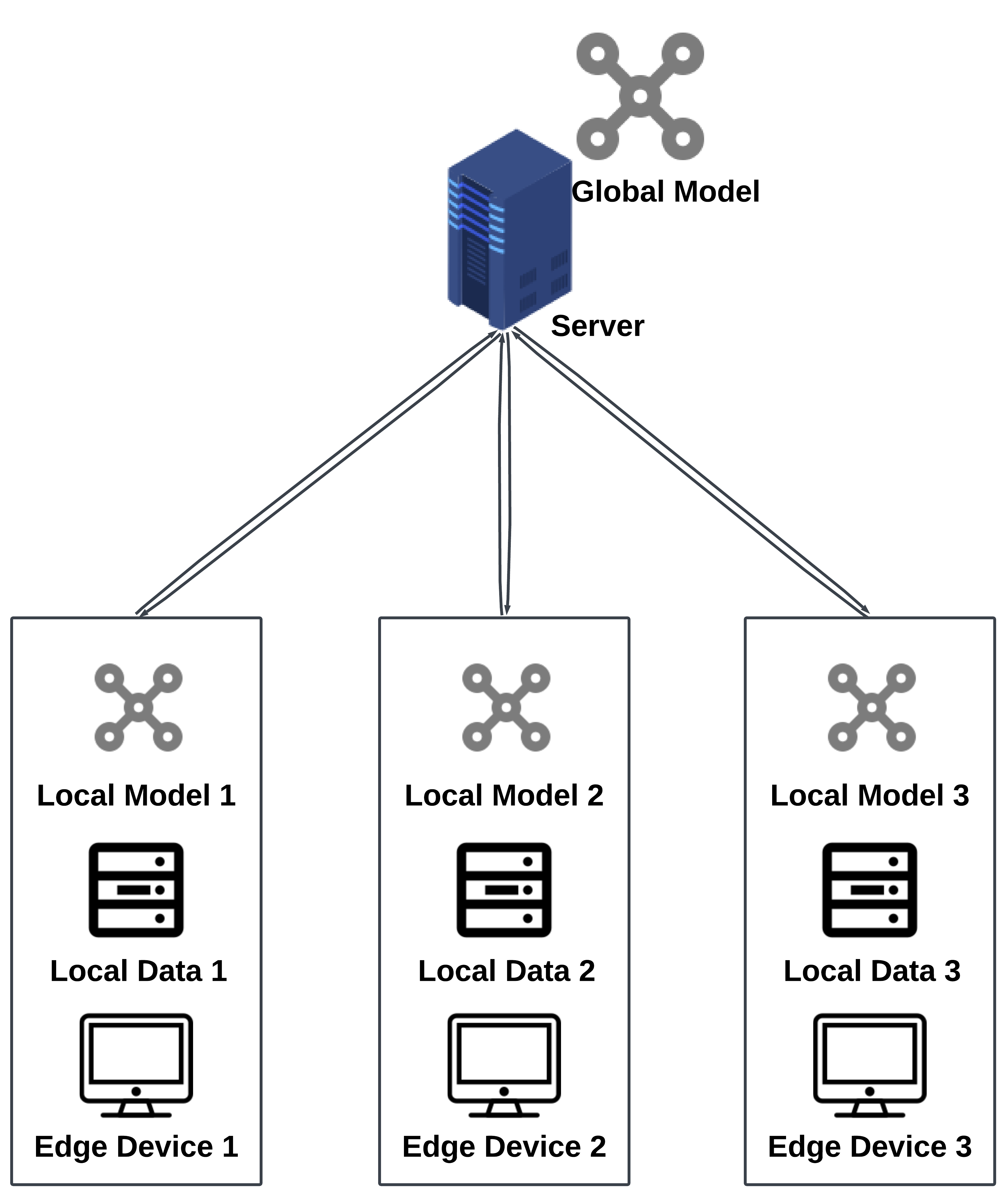}}\hspace{10 mm}
    \subfloat[]{\includegraphics[width=0.349\linewidth]{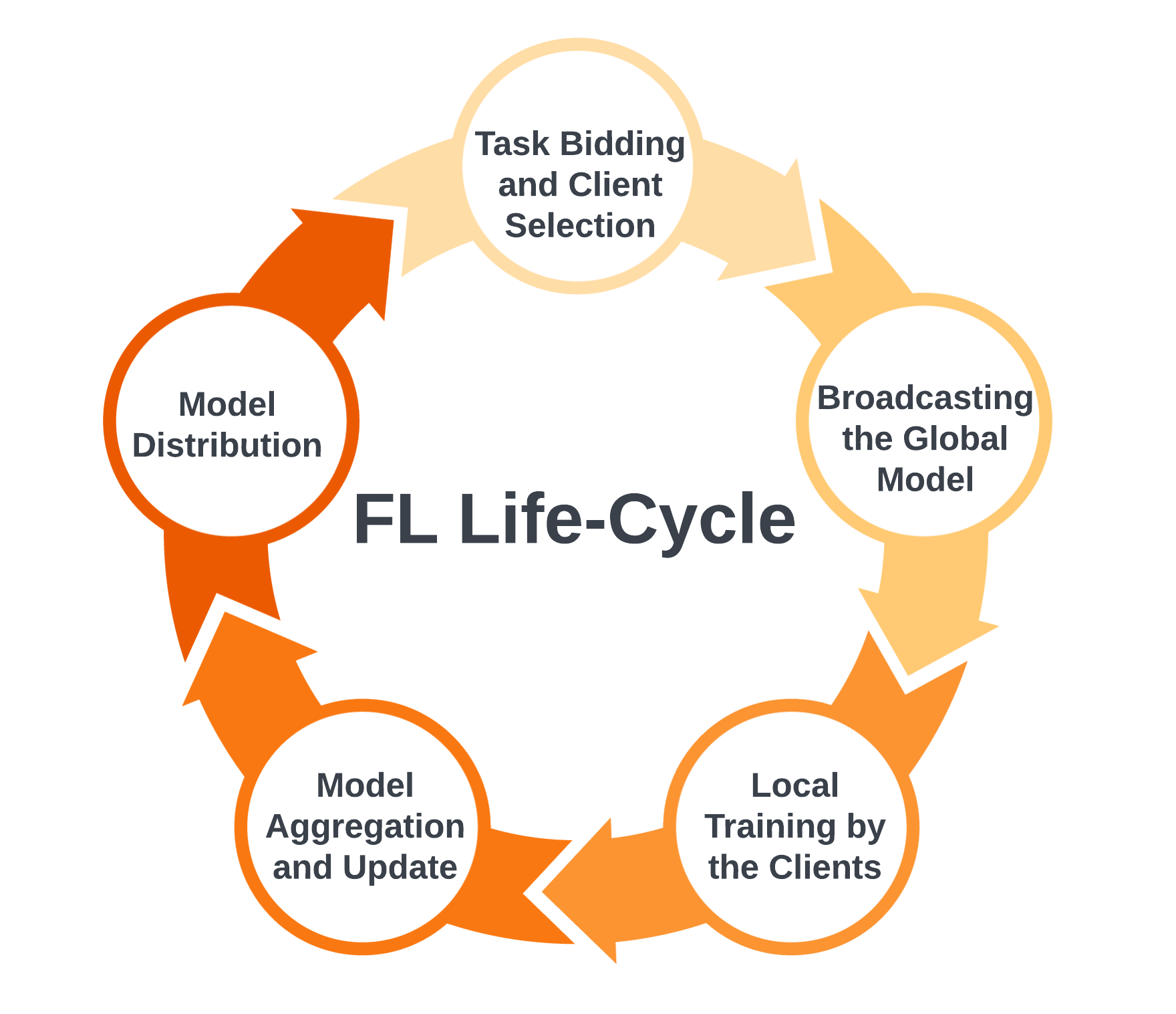}}
    \caption{(a) Federated Learning Model Aggregation: Shows how local model updates from multiple clients are combined on the server using aggregation methods to form a global model without sharing raw data. (b) FL Life Cycle}
    \label{fig:Federated_Learning}
\end{figure*}

The federated learning life cycle consists of 6 stages \cite{li2024threats} (see Fig.~\ref{fig:Federated_Learning}(b)) as follows:
\begin{enumerate}
    \item Task Bidding and Client Selection: It begins with service providers announcing a learning task within an FL community. Interested clients submit bids to participate, offering details about their available resources, such as computing power, bandwidth, network capabilities, and training data. The service provider then reviews these bids and selects the clients that best meet the task’s requirements.
    \item Broadcasting the Global Model: Once clients are chosen, the service provider or a third-party server transfers the current global model to them. During the first round of training, the server also sets the conditions for the end of both local and global learning iterations, ensuring that participants know the expected milestones.
    \item Local Training by the Clients: After receiving the global model, each participant begins training the model locally using their local data. This process continues over several rounds, with the clients working towards a set target, like a specific model accuracy or a fixed number of training rounds. Once local training is complete, clients send their updated models back to the server.
    
    \item Model Aggregation and Update: The server aggregates all the received local updates using a predefined method such as FedAvg (Federated Averaging) \cite{li2019convergence}, and this combined update is used to adjust the global model. 

    \item Model Distribution: Finally, the updated global model is distributed back to the clients for further training.
\end{enumerate}

\section{Related Works}

In this section, we discuss existing research papers on FL and explain how this survey paper advances beyond them.  Yang et al. \cite{yang2019federated} summarized FL, covering the foundational concepts and applications of FL. The paper discusses FL architectures, privacy preservation techniques, and future directions in the field. However, it lacks a detailed discussion on deployment scenarios, communication methods, and aggregation strategies, which are crucial for the practical implementation of FL in real-world scenarios. 

The interconnectedness of IoT networks raises significant security concerns due to the sensitive data recorded by various edge nodes. Aledhari et al. \cite{9153560} summarized FL technologies, protocols, and applications. The paper delves deeply into FL architectures, various frameworks, network protocols, and the applications of FL, including its use in IDS. However, it lacks in-depth discussion on various aggregation techniques and deployment scenarios and advanced privacy-preserving techniques, which are vital for practical and secure implementations of FL in real-world scenarios. Zhang et al. \cite{zhang2021survey} presented a comprehensive survey on FL. The paper introduced the existing work of federated learning from five aspects: Data Partitioning, Privacy Mechanism, Machine Learning Model, Communication Architecture, and Systems Heterogeneity. The survey offers an in-depth analysis of these aspects, providing foundational insights into the technical and theoretical aspects of FL. However, the paper has notable limitations. It does not explore the applicability of FL in IDS, a critical area where privacy-preserving methods could have a significant impact given the sensitivity of network traffic data. Additionally, the survey lacks a detailed discussion on FL deployment scenarios, a crucial aspect for practical implementation in real-world systems. Another gap is the limited coverage of diverse model aggregation techniques, which plays a central role in achieving effective global models while addressing challenges such as communication overhead and heterogeneity among participating clients. Nguyen et al. \cite{nguyen2021federated} reviewed the application of FL in IoT networks, emphasizing how FL can preserve the privacy of sensitive data while enabling collaborative learning. The survey extensively discusses the applications of FL in IoT, its network structures, and architectures, providing valuable insights into the intersection of FL and IoT. However, the survey lacks a detailed exploration of deployment scenarios, which are critical for real-world implementation, as well as an in-depth analysis of model aggregation techniques that address the challenges of data heterogeneity and communication efficiency. Additionally, the absence of a discussion on standardized frameworks for implementing FL in IoT networks limits its utility for practitioners aiming for scalable and robust deployments. Mothukuri et al. \cite{mothukuri2021survey} presented an overview of the security and privacy aspects of FL. The paper provides an in-depth overview of FL, including various aggregation techniques and architectures, and a detailed classification of security threats and attacks within the FL domain. Additionally, it extensively discusses defensive techniques to mitigate these vulnerabilities, making it a valuable resource for researchers exploring security-centric FL methodologies. However, the paper is primarily focused on security and privacy concerns and does not cover network protocols in detail, which are essential for addressing communication challenges in FL systems.

In a world where data privacy and collaborative intelligence converge, FL emerges as a transformative solution. Abdulrahman et al. \cite{9220780} discussed the transition from centralized approaches to FL. The paper provides a detailed discussion of the privacy and security concerns and challenges associated with FL, its applications, and the optimization of federated algorithms. However, the paper lacks a focused survey on the use of FL for IDS and an exploration of deployment scenarios in practical applications. It also does not mention deep learning using FL. Li et al. \cite{li2021survey} discussed FL, providing an overview of its fundamental concepts, including various FL architectures, different machine learning models in FL, deployment scenarios, and communication architectures. However, the paper lacks a detailed discussion on various aggregation techniques and the application of FL in IDS. Banabilah et al. \cite{banabilah2022federated} presented a survey on FL, discussing various FL architectures, their applications, and the challenges associated with their implementation. However, the paper lacks a detailed discussion on deployment strategies, various aggregation algorithms, and the specific applications of FL in IDS. Wen et al. \cite{wen2023survey} presented a survey on FL that provides foundational knowledge on the subject. The paper addresses key aspects of FL, including privacy and security protection mechanisms, communication overhead challenges, and heterogeneity issues. Furthermore, it highlights various applications of FL, including its use in IDS. However, the survey falls short in several critical areas. It lacks detailed discussions on standardized frameworks, effective communication techniques, and practical deployment scenarios for FL. Moreover, the paper does not provide an in-depth analysis of the diverse model aggregation techniques crucial for optimizing global models, particularly in environments characterized by high data heterogeneity and constrained resources. Khraisat et al. \cite{khraisat2024survey} provided a comprehensive overview of FL, addressing key aspects such as FL architectures, communication methods, aggregation techniques, and FL frameworks. However, the study does not delve into federated deep learning specifically for IDS and lacks coverage of network protocols.

None of the existing surveys discuss quantum federated learning, particularly focusing on network security. Most of the surveys also overlooked model compression techniques and communication methods, which are one of the most important aspects in resource-constrained distributed devices.

\begin{table*}[ht]
\centering
\caption{Comparison of Existing Survey Papers on FL}
\renewcommand{\arraystretch}{1} 
\setlength{\tabcolsep}{1pt} 
\resizebox{\textwidth}{!}{
\begin{tabular}{@{}l*{11}{>{\centering\arraybackslash}p{2cm}}@{}}
\toprule
\textbf{Feature} 
& \scriptsize \textbf{\makecell{Yang\\et al.\\(2019) \\ \cite{yang2019federated}}}
& \scriptsize \textbf{\makecell{Aledhari\\et al.\\(2020) \\ \cite{9153560}}} 
& \scriptsize \textbf{\makecell{Zhang\\et al.\\(2021) \\ \cite{zhang2021survey}}} 
& \scriptsize \textbf{\makecell{Nguyen\\et al.\\(2021) \\ \cite{nguyen2021federated}}} 
& \scriptsize \textbf{\makecell{Mothukuri\\et al.\\(2021) \\ \cite{mothukuri2021survey}}} 
& \scriptsize \textbf{\makecell{Abdulrahman\\et al.\\(2021) \\ \cite{9220780}}} 
& \scriptsize \textbf{\makecell{Li\\et al.\\(2021) \\ \cite{li2021survey}}} 
& \scriptsize \textbf{\makecell{Banabilah\\et al.\\(2022) \\ \cite{banabilah2022federated}}} 
& \scriptsize \textbf{\makecell{Wen\\et al.\\(2023) \\ \cite{wen2023survey}}} 
& \scriptsize \textbf{\makecell{Khraisat\\et al.\\(2024) \\ \cite{khraisat2024survey}}} 
& \scriptsize \textbf{\makecell{This\\paper}} \\ \midrule

FL Architectures         & \checkmark & \checkmark & \checkmark & \checkmark & \checkmark & \checkmark & \checkmark & \checkmark & \checkmark & \checkmark & \checkmark \\
Aggregation Techniques   &            &            &            &  & \checkmark  & \checkmark           &            &            &            & \checkmark & \checkmark \\
Privacy Preservation     & \checkmark  &  & \checkmark & \checkmark  & \checkmark & \checkmark & \checkmark & \checkmark & \checkmark & \checkmark  & \checkmark \\
Deployment Scenarios    &            &            &            &            & \checkmark           &            &            &            &  &  \checkmark & \checkmark \\
Network Protocols    &            &   \checkmark        &            &            &         &            &            &            &  &  & \checkmark \\
Attack Specific Solutions                    &            &           &  &  &            &            &            &            &            &   & \checkmark \\ 
Model Compression/Optimization                    &            &            &   &            &            &            &            &           & \checkmark &  & \checkmark  \\
Quantum-Federated Learning                    &            &            &   &            &            &            &            &            & &  & \checkmark  \\

Communication Methods    &           &           &  \checkmark          &  \checkmark          &           &            &      \checkmark      &            &  & \checkmark & \checkmark \\
FL Framework                      &            & \checkmark           &            &  &  \checkmark          & \checkmark           &   \checkmark         &            &            &  \checkmark & \checkmark \\
QFL Framework                      &            &            &            &  &          &            &         &            &            &  & \checkmark \\
Intrusion Detection                      &            &            &  &  &            &            &            &            &   \checkmark         &  \checkmark & \checkmark \\ 
Future Directions              & \checkmark & \checkmark & \checkmark & \checkmark & \checkmark & \checkmark & \checkmark & \checkmark & \checkmark & \checkmark & \checkmark \\ \bottomrule
\end{tabular}
}
\label{tab:comparison}
\end{table*}

Table~\ref{tab:comparison} provides a comprehensive overview of how this paper differs from other existing survey papers. It highlights the unique contributions and broader scope of this work, addressing gaps left by previous studies. Notably, this paper stands out as the most up-to-date survey, encompassing recent advancements and covering aspects that earlier works have overlooked.

\section{Federated Learning in IDS}

\begin{figure*}
    \centering
    \includegraphics[width=1.00\linewidth]{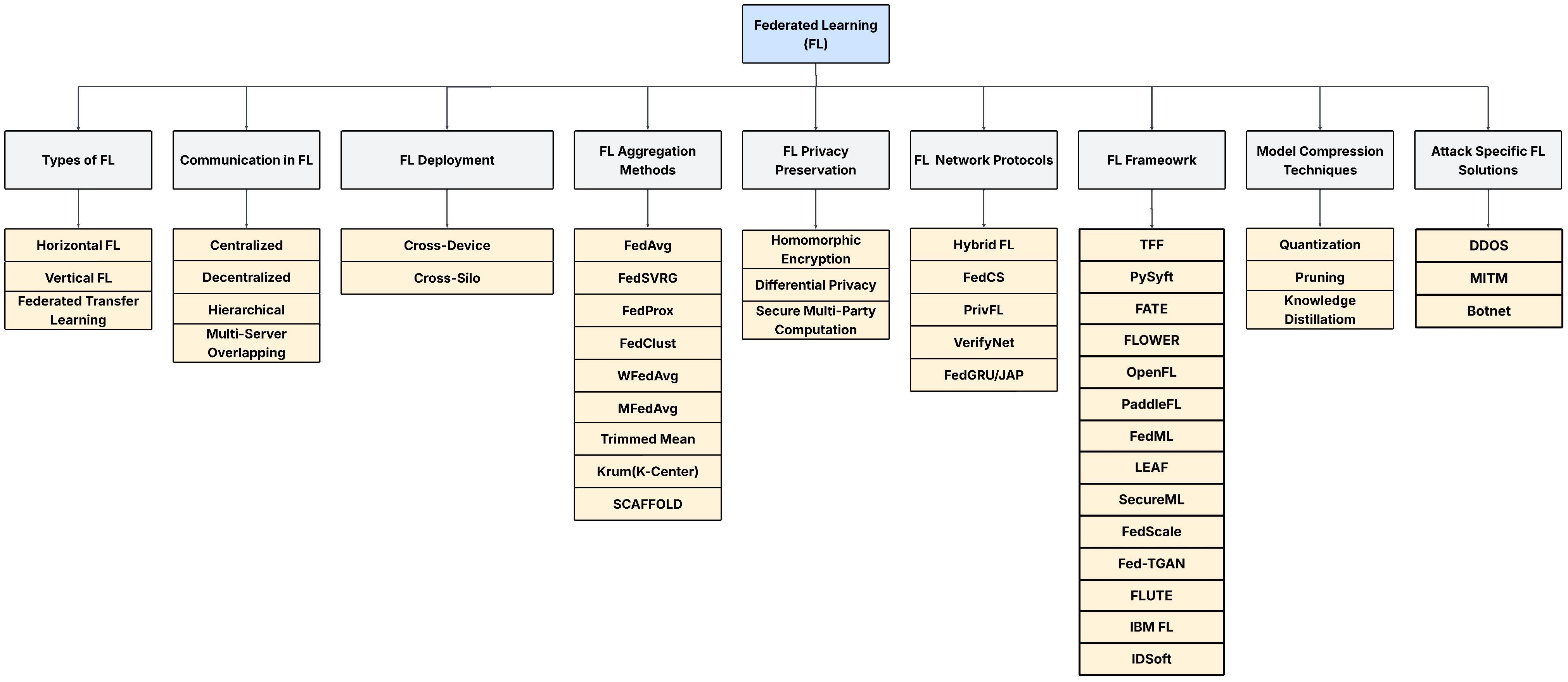}
    \caption{Tree diagram illustrating the taxonomy of Federated Learning in IDS, highlighting major categories and their subcomponents.}
    \label{fig:taxonomy}
\end{figure*}

Fig.~\ref{fig:taxonomy} presents a high-level view of Federated Learning in IDS, with its main categories and components organized hierarchically.

\subsection{Types of Federated Learning}
\begin{figure*}[ht]
        \centering
        \subfloat[]{\includegraphics[width=0.30\linewidth]{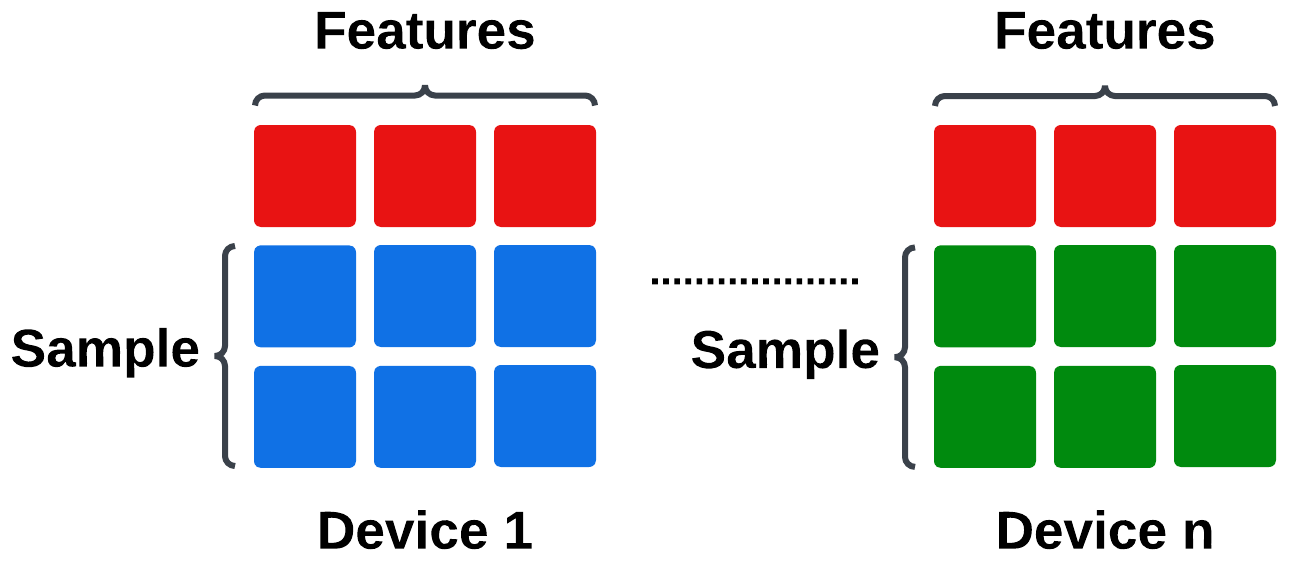}}
        \subfloat[]{\includegraphics[width=0.30\linewidth]{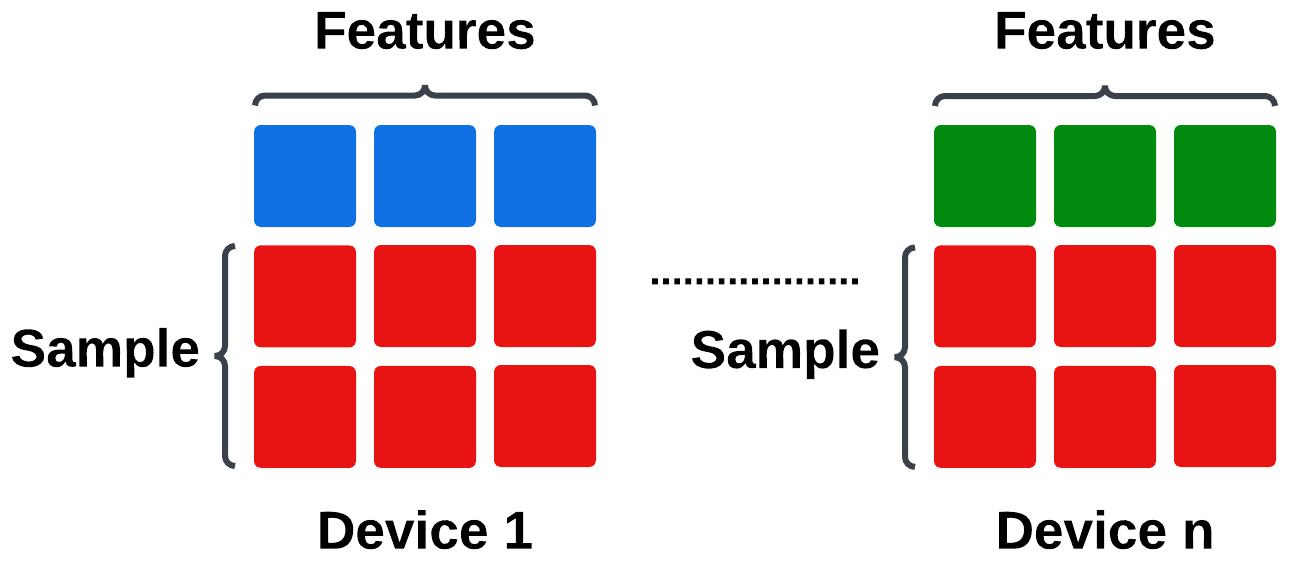}}
        \subfloat[]{\includegraphics[width=0.30\linewidth]{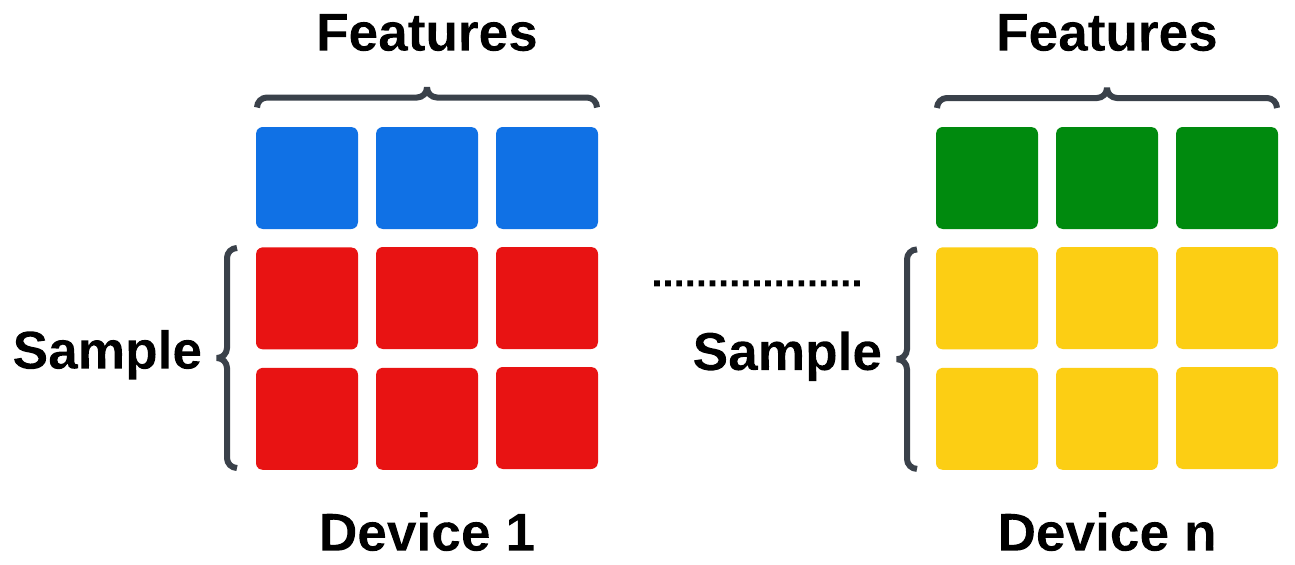}}
        \caption{Types of FL: (a) Horizontal FL (HFL) with shared features but distinct samples; (b) Vertical FL (VFL) with shared samples but distinct features; (c) Federated Transfer Learning (FTL) with distinct samples and features.}

        \label{fig:Types of Federated Learning}
        
    \end{figure*}

FL can be categorized into various types depending on how the data is distributed and how the model is trained across the participants. The primary types of FL are Horizontal Federated Learning (HFL), Vertical Federated Learning (VFL), and Federated Transfer Learning (FTL) (see Fig. \ref{fig:Types of Federated Learning} )\cite{10.1145/3511285.3511291}.

\textbf{Horizontal Federated Learning (HFL)}: In Horizontal Federated Learning, the clients(devices) have the same set of features (attributes) but different samples (data instances)(see Fig. \ref{fig:Types of Federated Learning}(a)). This is typically the case when each client has data that represents a subset of the overall population, but the data is homogeneous in terms of features. Example: A smartphone app that collects health data, such as heart rate and steps taken. Different users’ phones generate records with the same set of features but different data instances \cite{nguyen2022federated}.

\textbf{Vertical Federated Learning (VFL):} In Vertical Federated Learning, the clients have different features(attributes) for the same set of samples(data instances)(see Fig. \ref{fig:Types of Federated Learning}(b)). This is typically the case when multiple clients have data about the same entities but with different attributes or features. Example: Two hospitals may have data on the same patients(same data instances), but one hospital has medical history data(e.g., diagnoses, treatments), while the other has lab results(e.g., blood tests, imaging reports) \cite{liu2024vertical}.

\textbf{Federated Transfer Learning (FTL):} Federated Transfer Learning (FTL) combines the principles of both Federated Learning and Transfer Learning (see Fig. \ref{fig:Types of Federated Learning}(c)). In FTL, a model is trained across a set of clients with limited data or resources in such a way that knowledge from related domains or tasks is transferred to improve the learning process in a target domain. It is especially useful in situations where one client has insufficient data to train a model effectively but can leverage knowledge from other clients or previously trained models. Example: A model trained on one set of medical data(e.g., detecting pneumonia from X-ray images) can be fine-tuned with a smaller dataset from a different hospital to detect other types of medical conditions, such as tuberculosis, without requiring extensive data from the new hospital \cite{chen2020fedhealth}.

\subsection{Communication Methods}

\begin{figure}[ht]
    \centering
    \subfloat[]{\includegraphics[width=0.24\textwidth]{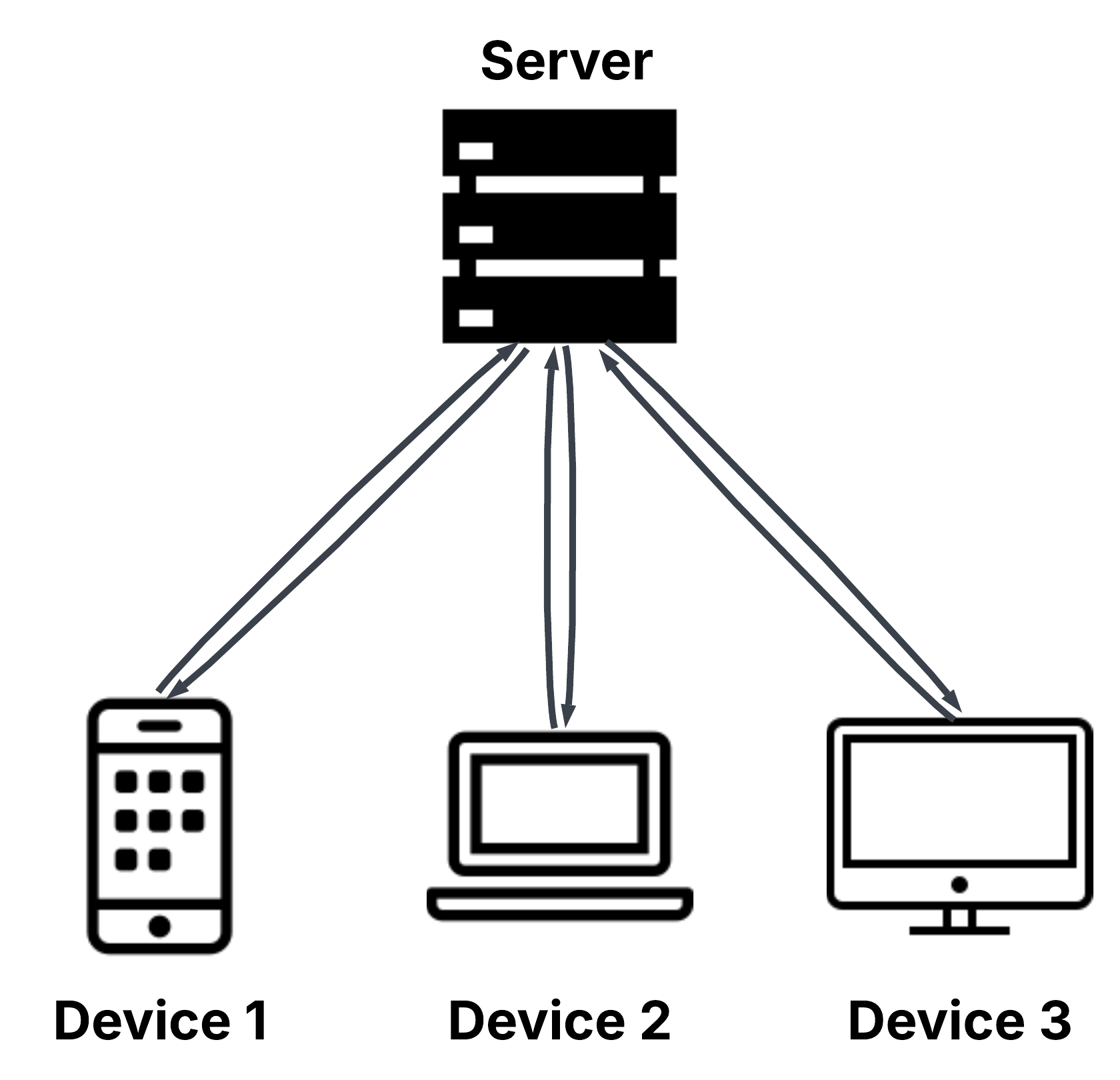}}
    \hfill
    \subfloat[]{\includegraphics[width=0.24\textwidth]{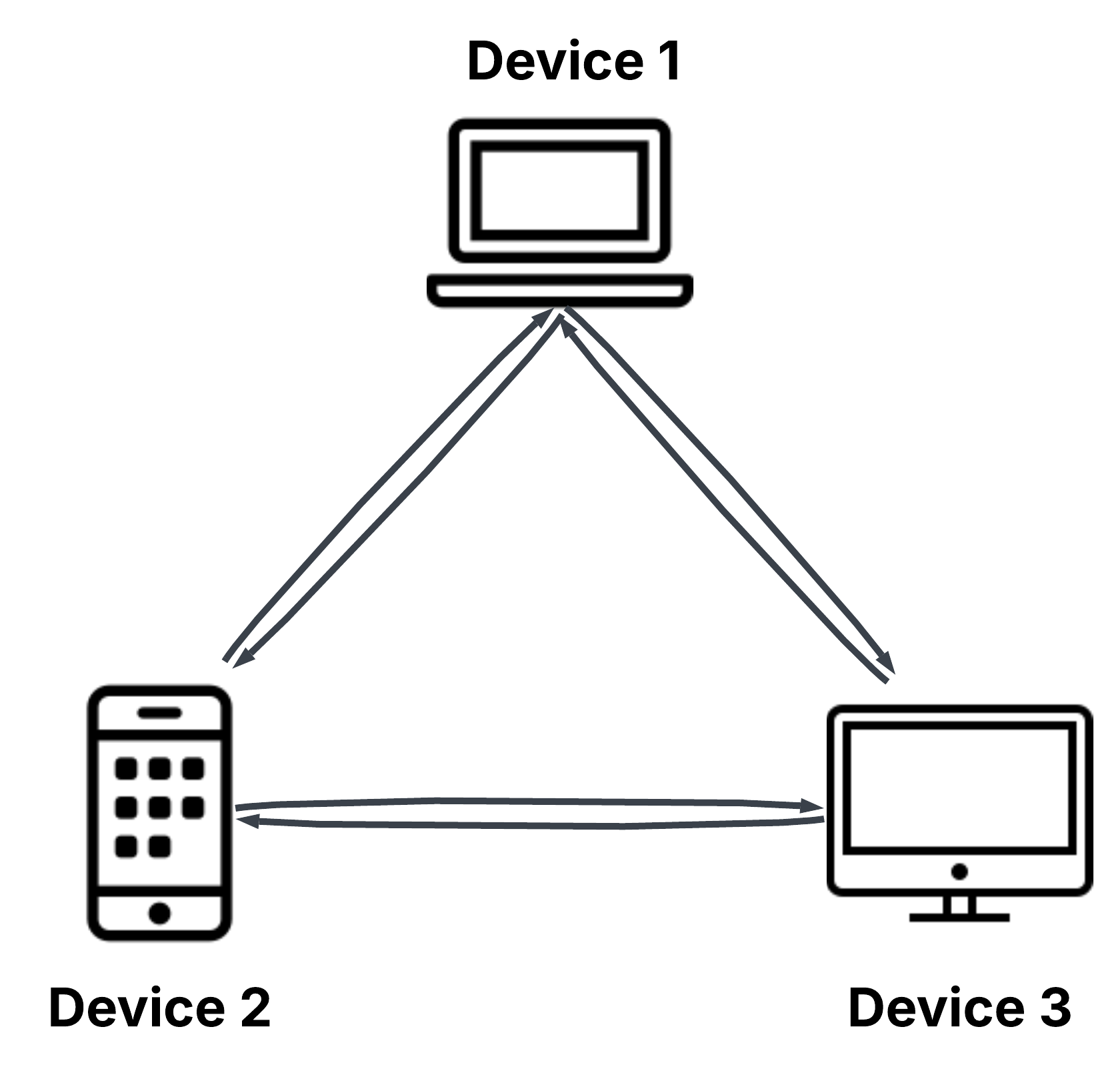}}
    \hfill
    \subfloat[]{\includegraphics[width=0.24\textwidth]{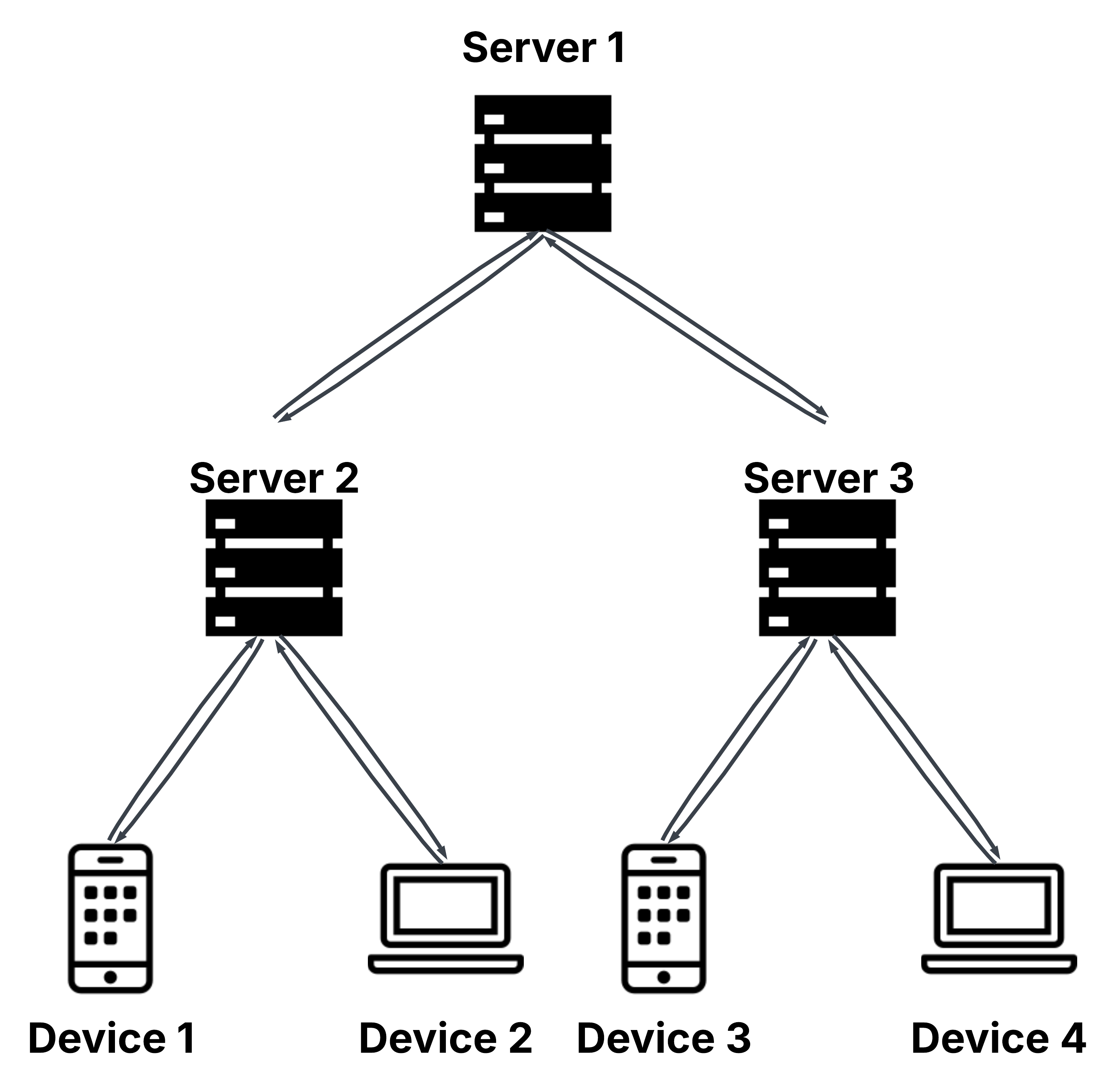}}
    \hfill
    \subfloat[]{\includegraphics[width=0.24\textwidth]{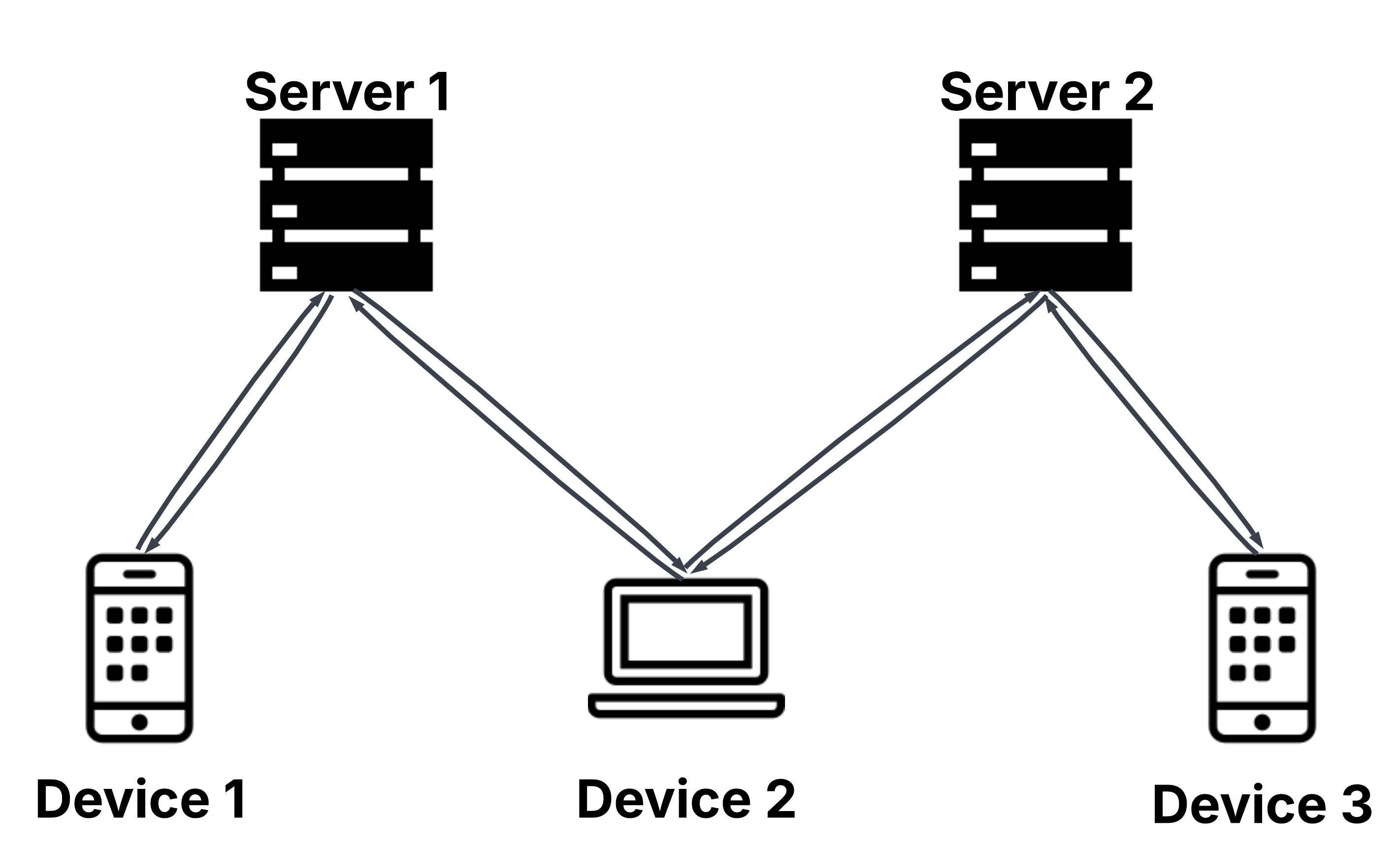}}
    \hfill
    \caption{Communication Methods: (a) Centralized Federated Learning(Client-Server Architecture); (b) Decentralized Federated Learning (Peer-to-Peer); (c) Hierarchical Federated Learning (Multi-Tier Architecture); (d) Multi-Server Overlapping}
    \label{fig:commuication}
\end{figure}

\textbf{Centralized Federated Learning (Client-Server Architecture):} In this type of communication, there is a central server to which all the clients are connected (see Fig.~\ref{fig:commuication} (a)). All the clients send their model updates to the server where the server aggregates the models received by the client to produce a global model which is then again distributed back to the clients \cite{khraisat2024survey}.

\textbf{Decentralized Federated Learning (Peer-to-Peer):} This communication does not have a dedicated server for model aggregation. Clients communicate directly with each other peer-to-peer to update their models (see Fig.~\ref{fig:commuication} (b)). In this setup, each client performs local training on its data, and after a certain number of training iterations, clients share their model updates with neighboring peers \cite{mothukuri2021survey,yuan2024decentralized}. The process continues iteratively across the network, with clients collaboratively improving the global model without relying on a centralized coordinator. 

\textbf{Hierarchical Federated Learning (Multi-Tier Architecture):} It organizes the communication process in multiple layers, where clients in the first layer (often edge devices or local nodes) train their models on local data and then sends their model updates to higher layers for aggregation \cite{khraisat2024survey} (see Fig.~\ref{fig:commuication} (c)). In this setup, the first layer of clients (typically referred to as leaf nodes) performs local training and sends their updates to the second layer, which consists of aggregators or intermediate nodes (such as edge servers or routers). These aggregators perform partial global aggregation of the updates received from the first layer, combining them into a more refined model. This aggregated model is then sent to the next layer for further aggregation or to a central server for the final global model.

\textbf{Multi-Server Overlapping:} This communication technique is used in distributed and federated learning systems, where multiple servers are involved in the communication and aggregation process \cite{khraisat2024survey} (see Fig.~\ref{fig:commuication} (d)). In this approach, several servers, often located in different regions or networks, handle different parts of the communication and model aggregation tasks. These servers overlap in terms of the data or model updates they handle, working collaboratively rather than independently.

\subsection{FL Deployment}

\begin{figure*}[t]
        \centering
        \subfloat[]{\includegraphics[width=0.40\linewidth]{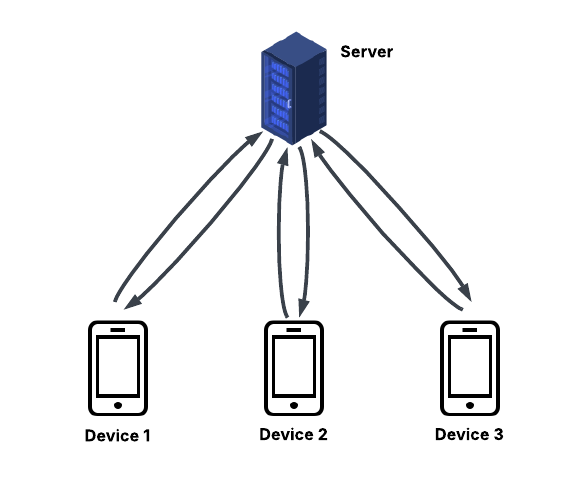}}
        \subfloat[]{\includegraphics[width=0.40\linewidth]{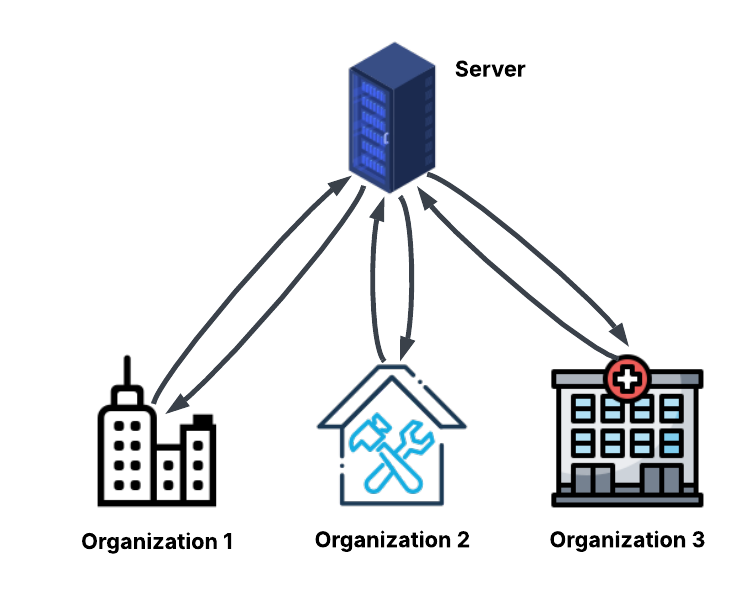}}

        \caption{FL Deployment Methods: (a) Cross-Device FL Deployment; (b) Cross-Silo FL Deployment.}

        \label{fig:fl deployment}
        
    \end{figure*}

FL deployment can be categorized into cross-device and cross-silo deployment \cite{khraisat2024survey} (see Fig.~\ref{fig:fl deployment}). 

\textbf{Cross-Device Federated Learning:} It involves a large number of edge devices (e.g., smartphones, IoT devices) with diverse computational capabilities and huge connectivity. Edge devices operate on a massive scale with dynamic participation. It focuses on personalized model training and privacy preservation at the individual user level. The data distribution can be highly non-IID due to diverse user environments. Edge devices usually have resource constraints, such as limited energy and computational capabilities. The key challenges include scalability, energy efficiency and device heterogeneity.

\textbf{Cross-Silo Federated Learning:} It involves collaboration among a limited number of reliable entities (e.g., hospitals, financial institutions). It emphasizes secure and regulated model training across organizations while preserving data privacy. It has a small-to-medium scale with a fixed number of participants. It has relatively fewer resource constraints as compared to cross-device federated learning. The key challenges include regulatory compliance and inter-organizational trust.

\subsection{FL Model Aggregation}
\label{sec:model_aggregation}
\begin{figure}
    \centering
    \includegraphics[width=0.80\linewidth]{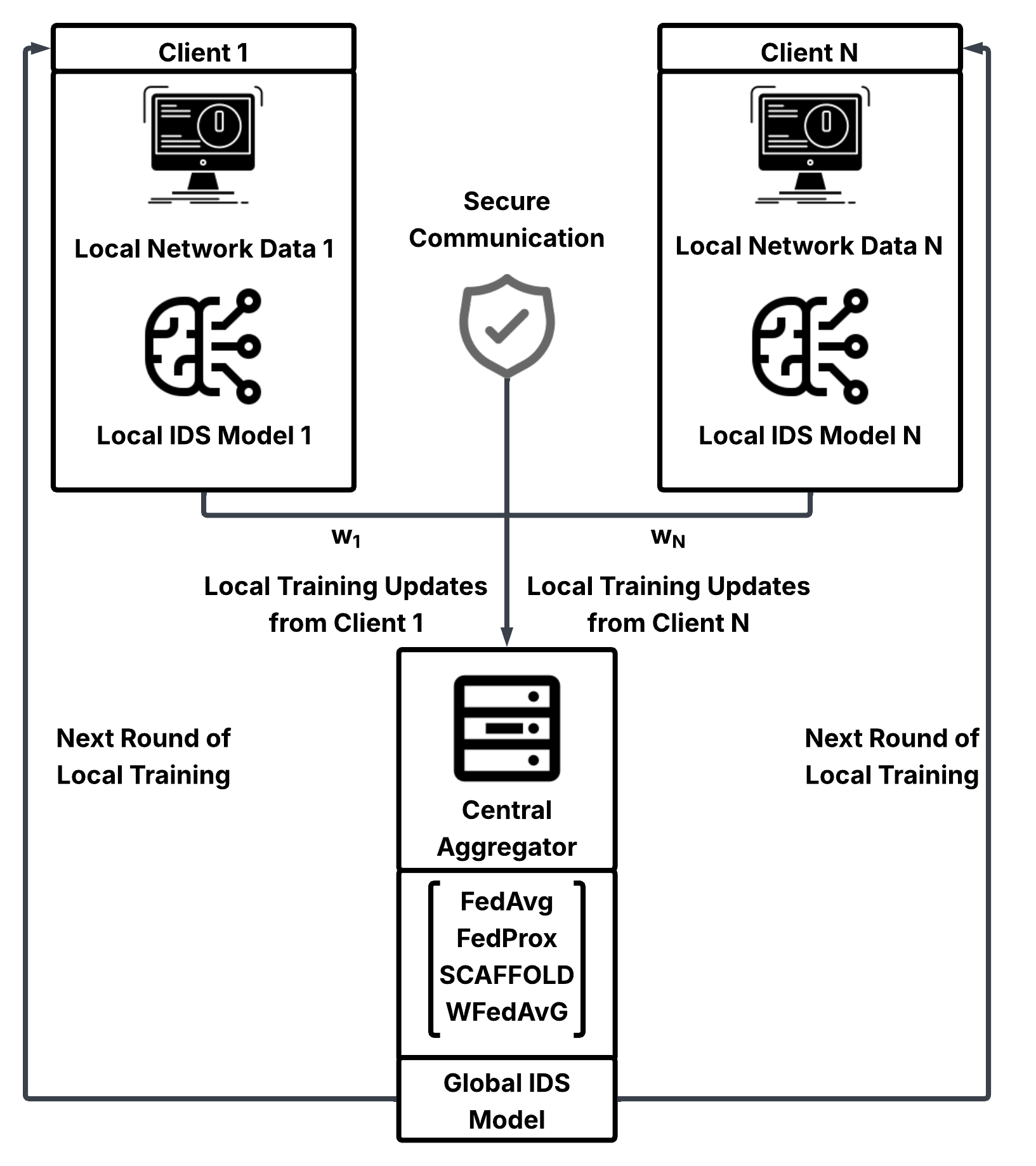}
    \caption{Federated Learning Model Aggregation: Shows how local model updates from multiple clients are combined on the server using aggregation methods to form a global model without sharing raw data.}
    \label{fig:aggregation}
\end{figure}

In FL, aggregation refers to the process of combining the model updates (usually weights or gradients) from multiple local devices to update the global model. Various aggregation techniques are used in Federated Learning methods \cite{qi2024model,khraisat2024survey, nanayakkara2024understanding}. This paper aims to provide a comprehensive overview of the aggregation methods, encompassing all known approaches based on our research and survey. Fig.~\ref{fig:aggregation} shows an overview of FL model aggregation.

\textbf{Federated Averaging (FedAvg):} It is the most common aggregation technique in federated learning. It aggregates model updates by averaging the local model weights from each client \cite{qi2024model}. The general formula for Federated Averaging  (FedAvg) is
\[ \mathbf{w}^{t+1} = \frac{1}{N} \sum_{i=1}^{N}  \mathbf{w}_i^t, \] 
where \( \mathbf{w}^{t+1} \) is the updated global model at round \( t+1 \), \( \mathbf{w}_i^t \) is the local model of the client \( i \) at round \( t \), \( N \) is the number of clients participating in the training at each round.

\textbf{Federated Stochastic Variance Reduced Gradient (FedSVRG):} It is a variant of FedAvg that uses variance reduction to make the aggregation more efficient, aiming to reduce the number of rounds needed for convergence \cite{li2021survey}. Each client computes a gradient approximation with reduced variance, and these approximations are aggregated similarly to FedAvg.

\textbf{Federated Proximal (FedProx):} This technique introduces a proximal term to the objective function, regularizing the local updates to prevent over-fitting to noisy data and reducing the effect of heterogeneous data across clients \cite{li2020federated}. Mathematically it can be represented as: 
\[w_{t+1} = \arg\min_{w} \left( \sum_{i=1}^{N} \frac{n_i}{n_{\text{total}}} L_i(w) + \frac{\mu}{2} \|w - w_t\|^2 \right),\]
where $w_{t+1}$: represents the updated model at time step $t+1$, $\arg\min_{w}:$ the minimization argument, $\sum_{i=1}^{N}:$ summation from $i = 1$ to N, $n_i/n_{\text{total}}:$ the weighting of the loss for client i, $L_i(w):$ the local loss function for client i, $\frac{\mu}{2}$ $\|w - w_t\|^2$: the regularization term with a constant $\mu$.

\textbf{Federated Learning with Clustering (FedClust):} Federated Clustering enables unsupervised learning in decentralized systems by collaboratively performing clustering across distributed clients while preserving data privacy. Each client \( k \) minimizes a local clustering objective, such as the k-means loss: 
\[ \mathcal{L}_k = \sum_{i \in D_k} \min_{c_j \in C_k} \|x_i - c_j\|^2, \]
where \( D_k \) is the dataset on client \( k \), \( C_k \) represents the local centroids, and \( c_j \) is the \( j \)-th centroid. The global server aggregates the local centroids \( C_k \) from \( K \) clients to update global centroids \( C \) using: $C = \frac{1}{K} \sum_{k=1}^K C_k$ \cite{islam2024fedclust}. This iterative process refines clustering results while ensuring that raw data remains on clients, maintaining privacy and reducing communication overhead.

\textbf{Weighted Federated Averaging (WFedAvg):} It is a variation of the Federated Averaging  (FedAvg) algorithm, where the contribution of each client to the global model is weighted based on the size of its local dataset \cite{khraisat2024survey}. This ensures that clients with larger datasets have a proportionally greater influence on the aggregated global model, addressing imbalances in data distribution across clients. Mathematically, the global model update in Weighted FedAvg can be expressed as: 
\[ w^{(t+1)} = \sum_{i=1}^N \frac{n_i}{n_{\text{total}}} w_i^{(t)}, \]
where \( w^{(t+1)} \): the updated global model at iteration \( t+1 \), \( w_i^{(t)} \): the model parameters from client \( i \) at iteration \( t \), \( n_i \): the number of data samples on client \( i \), \( n_{\text{total}} = \sum_{i=1}^N n_i \): the total number of samples across all clients.

\textbf{Median-Based Federated Averaging (MFedAvg):} MFedAvg is an algorithm that aggregates models using the median as a central measure. The global model update in Median Federated Averaging (MFedAvg) can be expressed as \cite{khraisat2024survey}:
\[ w^{(t+1)} = w^{(t)} + \text{Median}\left(\{\delta w_k^{(t)} \mid k \in S_t \}\right), \]
where \( \delta w_k^{(t)} = w_k^{(t)} - w^{(t)} \) represents the local updates from selected clients \( S_t \) at iteration \( t \).

\textbf{Trimmed Mean:} The global model is updated by aggregating the local model updates from clients while removing extreme outliers to ensure robustness \cite{khraisat2024survey,10200010}. The algorithm proceeds by first collecting local updates from selected clients, then sorting these updates and trimming a specified percentage of the extreme values from both ends. The remaining updates are then averaged to compute the aggregated update for the global model. Mathematically, it can be represented as: 
\[ w^{(t+1)} = w^{(t)} + \frac{1}{|K| - 2l} \sum_{k=l+1}^{|K|-l} \delta w_k^{(t)}, \]
where $w^{(t)}$: Global model parameters at iteration $t$, $w^{(t+1)}$: Global model parameters at iteration $t+1$, $\delta w_k^{(t)}$: Local model update from client $k$ at iteration $t$, $|K|$: Total number of connected clients, $l = \left\lfloor p \times |K| \right\rfloor$: Number of extreme updates to trim from both ends, where $p$ is the trim percentage, $\sum_{k=l+1}^{|K|-l}$: Summation over the remaining local updates after trimming the top and bottom $p \times |K|$ updates.

\textbf{Krum(K-Center):} In Krum Aggregation, the global model update is computed by selecting the local update from the client with the smallest distance-sum to its neighbors \cite{khraisat2024survey}. At each round, each client sends its local model update $\delta w_i$ to the server. The server computes the distance between every pair of updates $\delta w_i$ and $\delta w_j$ from clients $i$ and $j$, typically using Euclidean distance $d_{ij} = \| \delta w_i - \delta w_j \|$. For each client $i$, the server identifies its nearest neighbors $N_i$ by selecting the $n$-closest updates, where $n = p \cdot (|K| - 1)$, with $p$ being the trim percentage. The client $S_{\text{min}}$ with the smallest distance sum $Dis\_Sum(i) = \sum_{j \in N_i} d_{ij}$ is selected, and its update $\delta w_{S_{\text{min}}}$ is used to update the global model. Mathematically, the global model update is:
\[ w_{t+1} = w_t + \delta w_{S_{\text{min}}}, \]
where $S_{\text{min}} = \arg\min_i (\sum_{j \in N_i} \| \delta w_i - \delta w_j \|)$.

\textbf{Stochastic Controlled Averaging for Federated Learning (SCAFFOLD):} This algorithm improves federated learning by mitigating the effect of client data heterogeneity through the use of a control term \(c_i\) for each client \cite{praneeth2019scaffold}. At each round, a subset of clients is selected, and the global model \(x\) and control term \(c\) are sent to them. Each client updates its local model \(y_i\) based on the gradient of the loss function, adjusted by the difference between its control term \(c_i\) and the global control term \(c\), and computes a new control term \(c_i^{+}\). The updates \( \Delta y_i = y_i - x \) and \( \Delta c_i = c_i^{+} - c_i \) are then communicated back to the server. The server aggregates the updates by averaging them and applies global updates to both the model parameters and control terms, adjusting the global model \(x\) using \( \eta_g \Delta x \) and the global control term \(c\) using \( \frac{|S|}{N} \Delta c \), where \( |S| \) is the number of clients involved in the round and \(N\) is the total number of clients. This process helps to ensure more stable convergence in federated learning with non-i.i.d. data.

\subsection{FL Privacy Preservation}

The following section delves into the diverse privacy preservation techniques utilized in Federated Learning, highlighting their role in safeguarding data privacy while enabling collaborative model training \cite{mothukuri2021survey,kholod2020open,zhang2021survey}.

\textbf{Homomorphic Encryption:} It is a form of encryption \cite{rivest1978data} that allows computations to be performed on cipher texts (encrypted data) without the need to decrypt them first. This is powerful because it enables sensitive data to remain encrypted while being processed, ensuring privacy and security throughout the computation process \cite{khraisat2024survey}. The result of the computation, when decrypted, matches the result that would have been obtained if the operations had been performed on the plain text data. For example, in additive homomorphic encryption, $E(m1) + E(m2) = E(m1+m2)$, where $E(.)$ represents the encryption operation \cite{li2018cryptographic}.

\textbf{Differential Privacy:} It is a privacy-preserving technique that ensures individual data points in a dataset cannot be identified or inferred from the outputs of a data analysis or computation. The main idea is to add noise to the data or query results in a way that provides useful aggregate insights while protecting individual privacy \cite{khraisat2024survey}. One common method to achieve differential privacy is the Laplace mechanism, which adds noise drawn from the Laplace distribution to the output of a query. If \( \mathcal{A} \) is a query function (e.g., counting or summing data), the Laplace mechanism perturbs the result by adding Laplace noise scaled by \( \frac{\Delta f}{\epsilon} \), where \( \Delta f \) is the sensitivity of the function (the maximum possible change in the output caused by a single data point change) and is \cite{9714350}: 
\[ \mathcal{A}(D) + \text{Lap}\left(\frac{\Delta f}{\epsilon}\right). \]

\textbf{Secure Multiparty Computation:} It is a cryptographic technique that allows multiple parties to jointly compute a function over their private inputs without revealing those inputs to each other \cite{khraisat2024survey}. The core idea is to divide the computation into smaller parts and distribute them among the parties, where each party only performs its part of the computation and shares the results in such a way that the final outcome can be obtained, but no party learns anything about the other parties' private inputs \cite{li2020privacy}. Mathematically, let \( f(x_1, x_2, \dots, x_n) \) be the function to be computed, where \( x_i \) represents the private input of party \( i \). In SMPC, each party computes a share \( s_i \) of the input and sends it to the other parties. The final result \( f(x_1, x_2, \dots, x_n) \) is then computed through the secure aggregation of the shares, typically via secret sharing or garbled circuits. For example, using additive secret sharing, the inputs \( x_i \) are split into shares \( s_{ij} \) such that $x_i = \sum_j s_{ij}$, and the parties collaboratively compute the result by combining their shares while maintaining the privacy of each \( x_i \).

\subsection{FL Network Protocols}
Communication is one of the most important parts of FL. There can be various issues related to privacy and traffic flow \cite{choudhury2002anonymizing} while communicating in FL. This led to the development of various network protocols particularly focusing on FL. Network protocols are mainly focused on wireless communication which is necessary for various industries such as healthcare \cite{8859260}. In this paper, we review several network protocols and techniques designed to enhance the efficiency, privacy, and robustness of Federated Learning systems \cite{9153560}.

\textbf{Hybrid Federated Learning (Hybrid FL):} It integrates both centralized and decentralized communication models, aiming to maintain a balance between efficiency and model accuracy in distributed learning environments. In this hybrid framework, a combination of local data aggregation (decentralized) and global aggregation through a central server (centralized) is employed, enabling the system to adaptively optimize both data exchange and model convergence. The decentralized component facilitates communication among local clients (such as edge devices or IoT nodes) in a peer-to-peer fashion, reducing the reliance on a central server for frequent updates, which in turn lowers the communication overhead. On the other hand, the centralized model ensures global coordination, allowing for aggregation of local models and achieving high accuracy by leveraging the broader, collective knowledge of all participating clients.

\textbf{Federated Compressed Sensing (FedCS):} It integrates the principles of compressed sensing with federated learning to significantly reduce communication overhead in scenarios where network bandwidth is limited or constrained. In traditional federated learning, clients share their local model updates with a central server, which can result in substantial communication costs, especially when dealing with large datasets or high-dimensional data. FedCS addresses this challenge by applying compressive sensing techniques to compress the model updates before transmission, effectively reducing the amount of data sent over the network.

\textbf{Privacy-Preserving Federated Learning (PrivFL):} It is a framework designed to address privacy concerns in federated learning environments by employing advanced cryptographic techniques and privacy-enhancing methods to protect sensitive data during the entire learning process. Since federated learning relies on local data stored at client devices, ensuring that this sensitive data is not exposed during model training is critical, especially in applications such as healthcare, finance, and IoT, where data privacy is paramount.

\textbf{VerifyNet:} It introduces a set of verification mechanisms designed to ensure the integrity and authenticity of model updates in federated learning. It is particularly useful in networks where trust and data authenticity are critical, such as in sensitive environments or industries like healthcare, finance, and IoT. In federated learning, model updates are sent from multiple clients to a central server for aggregation. However, due to the distributed nature of the system, there is always a risk of malicious or faulty clients providing incorrect updates, which can compromise the overall model performance and reliability. It addresses this challenge by implementing verification protocols that authenticate and validate the model updates before they are aggregated. 

\textbf{FedGRU/JAP:} It is an approach that integrates Gated Recurrent Units (GRUs), a type of recurrent neural network (RNN), into federated learning frameworks for processing time-series data. GRUs are designed to capture temporal dependencies and long-term relationships in sequential data, making them highly effective for tasks such as anomaly detection, forecasting, and classification in dynamic environments. In federated settings, where data is decentralized and privacy concerns are crucial, FedGRU allows local devices to train RNN models on their own time-series data while sharing only aggregated updates, preserving privacy.

\subsection{FL Frameworks}
FL frameworks are software libraries or platforms developed to implement FL. These frameworks provide tools, APIs, and prebuilt functionalities to enable efficient communication, training, and aggregation across distributed clients in an FL setup \cite{kholod2020open}. The following frameworks have been widely adopted in research and industry for implementing FL-based methods:

\begin{itemize}
    \item \textbf{TensorFlow Federated (TFF):} It is an open-source FL framework developed by Google for experimenting and building FL algorithms \cite{TFF}. It comes with a built-in simulation environment that allows developers to simulate federated learning scenarios before deploying them to real devices. It has pre-built components for FL research, and customizable algorithms and allows integration with TensorFlow \cite{kholod2020open}.
    
    \item \textbf{PySyft:} It is an open-source library developed by OpenMined for secure and private machine learning, including FL \cite{ziller2021pysyft}. Its main focus is on privacy-preserving techniques in FL such as differential privacy and secure multi-party computation \cite{kholod2020open}. The core component of a PySyft library is the tensor which follows PyTorch or TensorFlow APIs but provides functionalities like remote execution and encryption. 
    
    \item \textbf{Federated AI Tech Enabler (FATE):} It is developed by WeBank and its primary focus is on industrial applications of FL \cite{FATE}. It supports different types of FL such as horizontal, vertical and transfer FL. It also has built-in modules for security \cite{kholod2020open}.
    
    \item \textbf{Flower:} It is a lightweight and modular FL framework for research and production \cite{beutel2020flower}. It has cross-framework compatibility (TensorFlow, PyTorch), supports edge devices and customizable strategies \cite{beutel2020flower}.
    
    \item \textbf{OpenFL:} It is developed by Intel and is an open-source FL framework for collaborative machine learning across multiple organizations \cite{reina2021openfl,OpenFL}. Its primary focus is on cross-silo FL \cite{mittone2024pushing}, and secure data collaboration. It also supports various ML frameworks.
    
    \item \textbf{PaddleFL:} This FL framework is based on PaddlePaddle by Baidu \cite{BaiduFL} with an stan 2.0 license. It supports horizontal and vertical FL, encryption and is designed for large-scale industrial applications \cite{kholod2020open}. This package includes FedAvg, SecAgg, and differentially private stochastic gradient descent (DPSGD).
    
    \item \textbf{FedML:} It is an open FL library for real-world FL and edge AI \cite{FedML}. It supports hierarchical FL, edge device integration, and benchmarking tools \cite{FedML}.
    
    
    \item \textbf{LEAF:} It is a benchmark framework designed for evaluating FL algorithms on realistic datasets and workloads \cite{caldas2018leaf}. Its primary focus is on non-IID data, customizable workloads, and extensibility \cite{caldas2018leaf}.
    
    \item \textbf{Secure ML:} It is a framework for privacy-preserving machine learning and federated learning \cite{7958569}. This framework implements secure computation techniques (e.g., homomorphic encryption, secure multiparty computation) \cite{7958569}.
    
    \item \textbf{FedScale:} It is an benchmarking FL framework that is designed to scale FL to large numbers of clients \cite{lai2022fedscale,FedScale}. FedScale primary focus is on scalability, optimized resource management, and dynamic scheduling of FL tasks \cite{lai2022fedscale}.
    
    \item \textbf{Fed-TGAN:} It is an FL framework for TGANs (Tabular Generative Adversarial Networks) that utilizes GANs for synthetic data generation in FL settings. It combines FL with GANs for data synthesis and improves FL for non-IID and limited datasets \cite{zhao2021fed}.

    \item \textbf{FLUTE (Federated Learning Utilities for Testing and Experimentation):} It is an open-source, high-performance framework developed by Microsoft Research for simulating federated learning scenarios at scale~\cite{garcia2022flute}.

    \item \textbf{IBM FL:} It is an open-source framework developed by IBM that enables collaborative machine learning across organizations without sharing raw data. It focuses on secure cross-silo training and supports multiple machine learning algorithms, including both classical and deep learning models~\cite{ludwig2020ibm}.
    
    \item \textbf{IDSoft:} An FL framework focusing on intelligent distributed systems and soft computing \cite{alotaibi2023idsoft}. It combines FL with distributed soft computing methods for resource optimization and adaptive learning \cite{alotaibi2023idsoft}.
    
\end{itemize}

Table~\ref{tab:fl_frameworks} gives a comparison between various open-source FL frameworks~\cite{riedel2024comparative}.
\renewcommand{\arraystretch}{1.3}
\begin{table*}[ht]
\centering
\caption{Comparison of Open-Source Federated Learning Frameworks}
\label{tab:fl_frameworks}
\begin{tabularx}{\textwidth}{l*{9}{>{\raggedright\arraybackslash}X}}
\toprule
\textbf{Feature / FL Framework} & \textbf{TFF} & \textbf{PySyft} & \textbf{FATE} & \textbf{Flower} & \textbf{OpenFL} & \textbf{PaddleFL} & \textbf{FedML} & \textbf{FLUTE} & \textbf{IBM FL} \\
\midrule
Security Mechanisms & Partial & Yes & Partial & Partial & Partial & Yes & Partial & Partial & Partial \\
FL Algorithms & FedAvg + adaptive & Manual (even FedAvg) & FedAvg only & FedAvg + adaptive & FedAvg + adaptive & FedAvg only & FedAvg + adaptive & FedAvg + adaptive & FedAvg + adaptive \\
ML Model Support & TensorFlow & TensorFlow \& PyTorch & TensorFlow \& PyTorch & TensorFlow \& PyTorch & TensorFlow \& PyTorch & PaddlePaddle & TensorFlow \& PyTorch & PyTorch & TensorFlow \& PyTorch \\
FL Paradigms & Horizontal & Horizontal \& Vertical & Horizontal \& Vertical & Horizontal \& Vertical & Horizontal & Horizontal \& Vertical & Horizontal \& Vertical & Horizontal & Horizontal \\
Edge Device Rollout & No & Yes (Raspberry Pi) & Partial & Yes & No & Partial & Yes & No & Yes \\
OS Support & MacOS & Windows \& MacOS & Linux & Windows \& MacOS & MacOS & Linux & Windows \& MacOS & Windows & Windows \& MacOS \\
GPU Support & Yes & Yes & Yes & Yes & Yes & Yes & Yes & Yes & Yes \\
Docker Support & Yes & Yes & Yes & Yes & Yes & Yes & Yes & No & Yes \\
Development Effort & Low & High & N.A. & Medium & Low & N.A. & Medium & Medium & Medium \\
Documentation / Tutorials & Extensive & Extensive & Partial & Extensive & Partial & Limited & Partial & Limited & Partial \\
Training Speed & Fast & Slow & N.A. & Medium & Fast & Medium & Medium & Medium & Slow \\
Data Preparation Effort & High & Low & N.A. & Low & Medium & High & Medium & Medium & High \\
Evaluation Methods & Built-in & Manual & N.A. & Partial & Built-in & N.A. & Partial & Built-in & Built-in \\
Pricing / Accessibility & Free & Free & Free & Free & Free & Free & Free & Azure ML & IBM Watson Studio \\
\bottomrule
\end{tabularx}
\end{table*}

\subsection{Model Compression Techniques for Communication Efficient Federated Learning}

    FL is used in various sectors such as healthcare, smart agriculture and finance \cite{kairouz2021advances}. One of the major challenges in FL is communicating model updates to the central server for aggregation. The complexity and size of the models can greatly impact the FL process. This highlights the need for efficient communication strategies \cite{le2022insights} such as various model compression techniques to reduce the communication overhead without compromising the model's performance.

    The most common types of model compression techniques include quantization, pruning, and knowledge distillation \cite{le2024survey}. Most of the existing work on FL uses full-precision weights that need to be transmitted to the server for aggregation. Quantization reduces memory and computation requirements by representing model parameters such as weights and activations using fewer bits compared to standard full-precision floating point numbers \cite{liu2021zero, cai2020zeroq}. This greatly reduces the size and computation requirements of the model without affecting the accuracy. Oh et al. in \cite{oh2022communication} and Oh et al. in \cite{oh2023fedvqcs} presented novel techniques FedQCS (Federated Learning via Quantized Compressed Sensing) and  FedVQCS (Federated Learning via Vector Quantized Compressed Sensing) respectively. In the FedQCS method, the gradients are compressed in the edge devices sequentially by block sparsification, dimension reduction and quantization. The compressed gradients are reconstructed in parameter server (PS) by minimum mean square error approach (MMSE) using expectation-maximization generalized-approximate-message-passing (EM-GAMP) algorithm \cite{vila2011expectation}. The convergence rate of the proposed method is given by \( O\left(\frac{1}{\sqrt{T}}\right) \), where T is the number of total iterations of the SGD algorithm. The results demonstrated that FedQCS with one bit overhead per gradient entry can attain a classification accuracy similar to the perfect reconstruction without compression. It outperformed other existing QCS-based frameworks both in terms of classification accuracy and normalized MSE of the gradient reconstruction. Similarly, in the FedVQCS method, the local model is compressed using dimensionality reduction followed by vector quantization. Gradients are reconstructed on the parameter server (PS) using a sparse signal recovery algorithm, in contrast to MMSE in \cite{oh2022communication}. The results showed that it achieved a 2.4\% improvement in accuracy compared to other state-of-the-art FL methods in the MNIST and FEMNIST datasets when the communication overhead is 0.1 bits per model entry.

    Most of the participating clients in the FL process involve devices having resource constraints such as limited power and computational and communication capabilities. The participating clients with weak capabilities can become the bottleneck of model training. To address these issues, model pruning is one of the effective ways. It reduces the redundant or less important parameters of a neural network which helps in reducing the size of the model, computational complexity and storage costs \cite{10330640}. Jiang et al. in \cite{jiang2023computation} developed Federated Model Pruning (FedMP) that simultaneously helps in reducing communication and computation overhead by adaptive model pruning over heterogeneous participating clients. A Multi-Armed Bandit (MAB) was implemented to adaptively determine pruning ratios for the heterogeneous participating clients. The parameter server (PS) determines different pruning ratios for different heterogeneous clients. A novel parameter synchronization scheme Residual Recovery Synchronous Parallel (R2SP) was also implemented to ensure the sub-models' parameters are properly synchronized before model aggregation. The proposed framework was implemented on a physical platform and the results showed that this method was efficient for various heterogeneous scenarios and can provide up to 4.1x speedup as compared to other existing FL solutions. Similarly, Yi et al. in \cite{yi2024fedpe} developed FedPE, a communication-efficient FL framework that allows each participating client to select personalized optimal local subnets for each round of FL. In each federated cycle, the server sends personalized subnets to each participating client. The clients test these subnets on their respective local data and compare the accuracies from previous round accuracies. If there are $k$ clients and $t$ represents each federated cycle, the difference between the accuracies at $t$ and $t+1$ is given by:
    \[
    \Delta \text{Acc}_k^{t+1} = \text{Acc}_k^{t+1} - \text{Acc}_k^t.
    \]
    Based on \( \Delta \text{Acc}_k^{t+1} \), the client decides whether to adjust the model size:

\begin{itemize}
    \item If \( \Delta \text{Acc}_k^{t+1} > 0 \) (accuracy improves), the client prunes the model to reduce its size.
    \item If \( \Delta \text{Acc}_k^{t+1} < 0 \) (accuracy drops), the client expands the model to increase capacity.
    \item If \( \Delta \text{Acc}_k^{t+1} = 0 \), no changes are made to the model.
\end{itemize}

    The pruning ratio depends on accuracy variation in local data. A strategy called error compensation strategy is implemented to select optimal subnets during pruning or expansion. After training, the clients upload the masks and remaining parameters to the server for aggregation. The pruned parameters are frozen to be 0 during local training. The server aggregates the local models using an accuracy-weighted fair aggregation rule and aligns the global model with each received local subnet. Finally, the server broadcasts the masks and remaining parameters back to the corresponding participating clients. The results showed that FedPE achieved 1.86x $-$121x improvement in communication efficiency with accuracy not being compromised. 

    Similarly, knowledge distillation is one of the methods of model compression in which knowledge is transferred from a larger mentor model to a smaller mentee model which reduces the communication overhead while maintaining high accuracy \cite{hong2022analysis, ge2024pfl}. Wu et al. in \cite{wu2022communication} proposed a novel technique called FedKD, a communication-efficient FL framework that communicates model parameters from a small mentee model in contrast to a large full-size model in standard FL solutions. It significantly reduces communication overhead while maintaining high accuracy. To maintain the performance of the mode, an adaptive mutual distillation strategy is introduced, where the mentor and mentee models learn reciprocally. Additionally, an adaptive loss weighting mechanism ensures high-quality knowledge transfer, and a Singular Value Decomposition (SVD)--based compression technique further reduces model update sizes. The results demonstrated a reduction in communication costs by 94.98\% outperforming other FL solutions in non-IID settings. Chen et al. in \cite{10090471} proposed a resource-aware transfer knowledge-based FL framework for resource-constrained devices. The proposed framework implements two grouping strategies, random and computing capacity grouping strategies. In random grouping, participating clients are grouped arbitrarily without the need for any private information whereas in computing capacity grouping clients are grouped on the basis for the computing capacity of the clients. Participating clients with limited computing capabilities are assigned lightweight models without overloading weaker devices. If the clients are participating in a first learning task, it either initializes the blank model, or in the case of subgroup training, the student model is adopted and the joint loss function is applied to incorporate knowledge distillation. The joint loss function is 
    \[
    F = \alpha F_{\text{su}} + (1 - \alpha) F_{\text{ta},}
    \]
    where, \( F_{\text{su}} \) and \( F_{\text{ta}} \) represent the loss function of the student model and the teacher model, respectively, and \( \alpha \) represents the hyperparameter used to control the weight of the teacher model. The results showed that the proposed framework was able to outperform other state-of-the-art schemes and substantially improve efficiency.

    Table~\ref{tab:model_compression} provides a comparison of common model compression techniques used in machine learning and federated learning for network security applications~\cite{dantas2024comprehensive}. 
It summarizes the working principle, key advantages, and potential limitations of \textbf{Quantization}, \textbf{Pruning}, and \textbf{Knowledge Distillation}, which are widely adopted to reduce model size, improve inference speed, and maintain performance in resource-constrained environments.

\begin{table*}[ht]
\centering
\caption{Comparison of Model Compression Techniques}
\label{tab:model_compression}
\begin{tabular}{lp{4cm}p{4cm}p{4cm}}
\toprule
\textbf{Technique} & \textbf{Description} & \textbf{Advantages} & \textbf{Limitations} \\ \midrule
Quantization & Reduces the precision of weights and activations (e.g., float32 $\rightarrow$ int8) & 
\begin{itemize} \item Lower memory footprint \item Faster inference \end{itemize} & 
\begin{itemize} \item Possible accuracy drop for very low precision \end{itemize} \\ \midrule
Pruning & Removes less important weights, neurons, or channels & 
\begin{itemize} \item Smaller model size \item Faster inference \end{itemize} & 
\begin{itemize} \item Requires careful retraining \item May reduce model capacity if over-pruned \end{itemize} \\ \midrule
Knowledge Distillation & Trains a smaller "student" model using a larger "teacher" model's outputs & 
\begin{itemize} \item Smaller model with similar performance \item Can improve generalization \end{itemize} & 
\begin{itemize} \item Needs pre-trained teacher model \item Additional training overhead \end{itemize} \\ \bottomrule
\end{tabular}
\end{table*}

\subsection{Attack Specific Federated Solutions}
\subsubsection{DDoS (Distributed Denial of Service) Attacks}
    A DDoS attack disrupts the services provided by the systems by flooding them with excessive requests, rendering them unable to function properly. A DDoS attack can have a severe impact on the federated learning process, slowing the model aggregation process or exhausting the limited resources of edge devices \cite{fotse2024federated}. Alhasawi and Alghamdi in \cite{alhasawi2024federated} presented a novel solution, Federated Learning for Decentralized DDoS Attack Detection (FL-DAD) using CNN, to detect DDoS attacks in IoT networks. They also mentioned the challenges while implementing their solutions which include achieving consistent accuracy due to heterogeneous data, resource constraints on edge devices, communication overhead, anomalous data intricacies and computation complexities. In spite of these challenges, FL-DAD constantly achieved an accuracy of 98\% for various DDoS classes, outperforming traditional centralized methods. 

    Similarly, Fotse et al. in \cite{fotse2024federated} developed FedLAD, to detect DDoS attacks in large-scale software-defined networks (SDN). XGBoost classifier was implemented to detect DDoS attacks within the SDN. They implemented three different aggregation strategies and compared the results for three different datasets. The results showed that Astraes and Ranking Client outperform FedAvg in most performance metrics and the results were almost similar to the centralized approach. This method helps in minimal resource consumption by distributing the detection process to multiple controllers enhancing detection speed. A few limitations of FedLAD are model accuracy being dependent on aggregation techniques and the necessity to protect the model against data poisoning threats.

\subsubsection{MITM (Man in the Middle) Attacks}

    MITM attacks are responsible for data theft, unauthorized access and system integrity damage. Various techniques such as ARP spoofing, session hijacking, DNS spoofing and SSL stripping are used for MITM attacks \cite{javeed2020man}. Satpathy et al. in \cite{satpathy2024enhancing} conducted a performance analysis using FL models on Gradient Boosting Machines (FL-GBM) and Long Short-Term Memory (FL-LSTM) combining them with PCA (Principle Component Analysis) for MITM attacks detection. The results showed that FL-GBM outperformed FL-LSTM in terms of accuracy and F1 score.

    Similarly, Pathmendre et al. in \cite{pathmendre2024codenexa} presented a novel security framework CodeNexa for MITM attack detection. Existing security mechanisms include hash functions, digital signatures and watermarking to protect the integrity of the data. These traditional methods though effective have several limitations. Hash functions verify static data integrity but are not able to adapt to the dynamic and evolving nature of FL \cite{shamir1979share}. Digital signatures maintain the authenticity of the data but the attacker can intercept and manipulate both data and signature during transmission \cite{rivest1978method}. Similarly, watermarks ensure ownership verification but fail to protect continuous model integrity \cite{samarati2001protecting}. The proposed method does not include cryptography; instead, it enhances model integrity by dynamic metric verification such as accuracy, precision, recall and area under the curve (AUC). During global model aggregation, the stored metrics are compared with newly calculated metrics from local updates before aggregating the model. The model update is accepted only if a certain threshold is met. The metrics need to be transmitted to the global server along with model weights which remains a challenge as it increases the communication overhead.

\subsubsection{Botnet Attacks}

    A botnet is a network of compromised devices connected to a network and is controlled remotely by a hacker (often called a botmaster) without the knowledge of device owners \cite{7842850}. Botnets pose a significant security threat to modern networks by enabling large-scale cyber attacks through coordinated actions by a network of compromised devices. Several works have explored FL for botnet detection. For instance, Kalakoti et al. in \cite{10679348} presented a unique way for botnet detection. In FL settings, achieving explainability poses a unique challenge as traditional XAI techniques are designed for centralized models and datasets. The proposed framework combined explainable AI with FL settings and was used for botnet detection. A deep neural network (DNN) and SHAP (SHapley Additive exPlanations) were used. SHAP is a post-hoc explainable method based on a game theory that helps in computing feature importance. SHAP was applied locally on each end device to get the local explainability of the model. The model parameters were aggregated using FedAvg along with the local SHAP explanations. To check the correctness, the aggregated SHAP explanation was compared with the global SHAP explanation generated by the centralized version of the model. Their results demonstrated a strong correlation between aggregated SHAP explanation and global SHAP explanation showing the effectiveness of XAI in FL settings. The results of the proposed framework showed that it was able to detect 99\% of the botnet in IoT networks along with explainability while maintaining data privacy. 
    
    Similarly, in \cite{10279423}, Zhang et al. proposed a new K-greedy aggregation algorithm based on uncertainty assessment to detect botnet attacks in an IoT network. The traditional FedAvg algorithm is most likely biased when new zero-day records are added to the detection model in the FL process. The proposed method involves two groups of two groups of IoT devices. Some unknown attacks (zero-day attacks) on one group are already present in another group's data. The models are trained locally on each IoT device. This is where Bayesian deep learning comes in. It helps each device to estimate how uncertain it is about its predictions. The K-greedy aggregation uses these uncertainty scores to decide which model to trust more during the aggregation process. The results show that it was able to detect zero-day attacks successfully.

    Table~\ref{tab:comparison-fed} shows an overall comparison of the papers discussed above.

\begin{table*}[htbp]
\caption{Comparison of Federated Learning Approaches for Network Intrusion Detection}
\centering
\begin{tabular}{p{2.5cm}p{1.2cm}p{2.5cm}p{1.5cm}p{2cm}p{2.7cm}p{1.3cm}}
\toprule
\textbf{References} & \textbf{FL Type} & \textbf{Aggregation Strategy} & \textbf{Deployment Method} & \textbf{Dataset} & \textbf{ML Model} & \textbf{Attack Type} \\
\midrule
Alhasawi et al.~\cite{alhasawi2024federated} & Horizontal & Weighted FedAvg & Cross-device & CICIDS2017 & CNN & DDoS \\
\midrule
Fotse et al.~\cite{fotse2024federated} & Horizontal & Astraes, Ranking Client, FedAVG & Cross-silo & InSDN, CICDDoS2019, CICDoS2017 & XGBoost & DDoS \\
\midrule
Satpathy et al.~\cite{satpathy2024enhancing} & Horizontal & FedAvg & Cross-device & UNSW-NB15 & GBM, LSTM with PCA & MITM \\
\midrule
Pathmendre et al.~\cite{pathmendre2024codenexa} & Horizontal & CodeNexa (validation-based) & Cross-device & MNIST & CNN & MITM \\
\midrule
Kalakoti et al.~\cite{10679348} & Horizontal & FedAvg & Cross-device & N-BaIoT & DNN & Botnet \\
\midrule
Zhang et al.~\cite{10279423} & Horizontal & K-greedy aggregation & Cross-device & N-BaIoT & Bayesian Deep Learning & Botnet \\
\bottomrule
\end{tabular}
\label{tab:comparison-fed}
\end{table*}

\section{Quantum Machine Learning (QML)}  
    Quantum computing is rapidly evolving and getting popular as it leverages the principles of quantum mechanics (i.e., superposition, entanglement, decoherence, and quantum interference) that process information in a different way as compared to classical computers \cite{hdaib2024quantum}. Unlike classical bits (i.e., 0 or 1), quantum bits (qubits) can exist in a superposition of both the states which helps in parallel computation across a vast solution space. A qubit can be in the state \( |\psi\rangle = |0\rangle \) or \( |\psi\rangle = |1\rangle \), and it can also be in a superposition state, i.e.,
\[
|\psi\rangle = \alpha|0\rangle + \beta|1\rangle \quad \text{with} \quad \alpha^2 + \beta^2 = 1.
\] 

The extreme values of the qubits are 0 and 1, but they can be anything in between, representing a state formed by the combination of 0 and 1. This is a superposition of two values. A qubit is capable of representing a larger amount of data since every state of a qubit can be mapped to different pieces of data. This makes a single qubit solve more computations as compared to many regular classical bits that work simultaneously. Quantum computers rely on quantum Turing machines for computation, in contrast to classical Turing machines in classical computers~\cite{deutsch1985quantum}. The runtime of an algorithm is measured in terms of the number of elementary operations $N$ for both classical and quantum computation~\cite{montanaro2016quantum}. In classical computing, the same problem can belong to NP but not P, meaning it can be verified in polynomial time but has no known polynomial-time solving algorithm on a classical machine, while it can belong to class BQP in quantum machines when defined on Hilbert spaces~\cite{corli2025quantum}. These distinctive features of quantum computing enable them to solve complex optimization and machine-learning problems involving high-dimensional data and intricate patterns. 

Quantum computing relies on Dirac notation (also called bra-ket notation) to efficiently describe and manipulate quantum states~\cite{dirac1981principles}. A quantum state is written as ket, denoted by $|\psi\rangle$. This represents a column vector in a complex vector space (Hilbert Space). The corresponding bra, denoted by $\langle \phi|$, is a conjugate transpose of a ket. It represents a row vector and is used to form inner products and matrix operations.

    \subsection{Quantum Gates and Circuits}

    In classical computing, the fundamental unit of classical information is bits (i.e., 0s and 1s). These bits can be manipulated using logic gates such as \textbf{AND}, \textbf{OR} and \textbf{NOT}, based on boolean algebra. For instance, a \textbf{NOT} gate flips 0 to 1 and vice-versa. These classical gates are the foundation of digital circuits and classical computing. These gates process bits by taking one or more bits as input and producing one or more output bits depending on the input.

    However, quantum gates operate on qubits that can exist in a superposition of 0 and 1. Quantum gates are reversible and linear. The output of these gates gives the probability of qubits being 0 or 1. The gates are classified based on the number of qubits they operate on. Gates that operate on a single qubit are called single-qubit gates, those on two qubits are called two-qubit gates, and those on three qubits are called three-qubit gates~\cite{steane1998quantum}. The most known single-qubit gates are Pauli X gate, Pauli Z gate and Hadamard (H) gate:

    X = $\left(\begin{array}{cc}
0 & 1 \\
1 & 0
\end{array}\right)$, Z = $\left(\begin{array}{cc}
1 & 0 \\
0 & -1
\end{array}\right)$, H = $\frac{1}{\sqrt{2}}\left(\begin{array}{cc}
1 & 1 \\
1 & -1
\end{array}\right)$

The Pauli X gate converts $|0\rangle$ to $|1\rangle$ and $|1\rangle$ to $|0\rangle$, similar to the NOT gate in digital circuits, the Pauli Z gate keeps $|0\rangle$ unchanged but converts $|1\rangle$ to $-|1\rangle$, and the Hadamard (H) gate converts $|0\rangle$ to $|+\rangle$ and $|1\rangle$ to $|-\rangle$.

    A quantum circuit is created by combining these gates to get desired outputs. The example of bell state~\cite{bell_state} shows how quantum gates can be used to create a quantum circuit and how quantum circuits can be used to perform various operations on quantum bits.

        Traditional quantum circuits are static - they are designed to perform fixed computations~\cite{10319321}. The concept of parameterized quantum circuits (PQCs), is introduced where certain gates in the circuit contain learnable parameters, similar to weights in classical neural networks. The PQCs are also called variational circuits. In QML, a PQC acts like a model, analogous to a neural network. Classical data are encoded into quantum states using encoding circuits or quantum feature maps~\cite{bhowmik2024quantum}. The variational circuit applies transformations based on trainable parameters on these quantum-encoded data. After measurement, classical values are used to calculate a loss function. Various classical optimizers (like gradient descent, Adam) can be used to adjust the parameters based on the loss function. The rotation angles and parameters in the quantum gates in a variational circuit are learned during QML. This learning occurs by minimizing the loss function, similar to classical settings.

    \subsection{Feature Encoding for Quantum Intrusion Detection}
    \label{feature_maps}

    The foremost step before implementing QML techniques is to map classical features into quantum states so that QML algorithms can operate on those features. Feature mapping is one of the most crucial steps in QML. Network traffic data used in IDS is high-dimensional, noisy and non-linearly correlated. The quantum feature maps exploit exponentially high-dimensional Hilbert space to encode classical features into quantum states~\cite{schuld2019quantum}, which allow quantum classifiers or kernels to detect subtle and complex patterns indicative of intrusions or attacks.

    Formally, a quantum feature map $\phi$ is a function that encodes the input vector $x$ into a quantum state $|\phi(x)\rangle \in \mathcal{H}$:

    \[
\Phi: \mathbb{R}^d \to \mathcal{H}, \quad x \mapsto |\phi(x)\rangle = U(x) |0\rangle^{\otimes n},
\]
where \( U(x) \) is a parameterized unitary operator that depends on the classical input \( x \), and \( n \) is the number of qubits used.

The common feature mapping techniques include~\cite{ibm_feature}:

\begin{enumerate}
    \item Angle Encoding (Rotation Encoding): It transforms each classical network features (such as packet size, duration, or entropy measures) into a rotation angle applied to a single qubit around a chosen axis: 
    \[
|\phi(x)\rangle = \bigotimes_{i=1}^d R_Y(x_i) |0\rangle = \bigotimes_{i=1}^d \left( \cos\frac{x_i}{2} |0\rangle + \sin\frac{x_i}{2} |1\rangle \right).
\]

Here, 
\[
R_Y(\theta) = e^{-i \theta Y / 2}
\]
denotes the rotation operator about the \( Y \)-axis by angle \(\theta\), where \( Y \) is the Pauli-\( Y \) matrix. This method maps each feature independently onto a qubit, making it hardware-friendly and straightforward to implement. One of the major challenges of angle encoding is that it requires one qubit per feature, which may become expensive for very high-dimensional IDS data.

    \item Amplitude Encoding: It leverages the amplitudes of a quantum state to represent the entire normalized network feature vector $\mathbf{x}$ compactly:

\[
|\phi(\mathbf{x})\rangle = \sum_{i=0}^{2^n - 1} x_i |i\rangle, 
\quad \text{with} \quad \|\mathbf{x}\|_2 = 1.
\]
Here, $n$ is the number of qubits, and $|i\rangle$ represents computational basis states. This encoding compresses $2^n$ classical features into $n$ qubits by exploiting quantum superposition. It enables compact representation of large feature vectors extracted from network flows or packet sequences, which is valuable for handling big data in security analytics. However, efficient quantum state preparation for arbitrary $x$ is resource-intensive and currently limits practical use on noisy quantum devices.

    \item Basis Encoding: It associates discrete or binary network features, such as protocol flags or categorical attributes, with computational basis states:
\[
|\phi(x)\rangle = |b(x)\rangle, \quad b(x) \in \{0,1\}^n.
\]

Each bit string $b(x)$ encodes the presence or absence of specific features or categories directly into the quantum state. It is naturally suited for categorical or binary IDS data (e.g., TCP flags, anomaly labels). However, it is less expressive for continuous-valued features like packet sizes or time intervals, which are common in network traffic.

    \item Phase Encoding: It represents each feature $x_i$ as a phase shift applied to a qubit:

\[
\lvert \phi(\mathbf{x}) \rangle 
= \bigotimes_{i=1}^{d} R_Z(x_i) \lvert + \rangle
= \bigotimes_{i=1}^{d} \frac{\lvert 0 \rangle + e^{i x_i} \lvert 1 \rangle}{\sqrt{2}},
\]

where 
\[
R_Z(\theta) = e^{-i \theta Z / 2},
\]
and $Z$ is the Pauli-$Z$ matrix.  

Unlike angle encoding, which affects measurement probabilities, phase encoding changes the relative phase between basis states. It is useful for time-dependent or periodic IDS features (e.g., traffic rate cycles, periodic scanning patterns) because phase shifts can naturally encode cyclic patterns and correlations.

    \item Dense Angle Encoding: It reuses qubits to encode multiple features by applying sequential rotations on the same qubit, often with different axes ($X$, $Y$, $Z$):  

\[
\ket{\phi(x)} = \prod_{j=1}^{m} \left[ \bigotimes_{i=1}^{n} R_{\alpha_{i,j}}(x_{i,j}) \right] \ket{0}^{\otimes n}.
\]

Here, $x_{i,j}$ is the $j$-th feature encoded on qubit $i$, $R_{\alpha_{i,j}}$ is a rotation gate around axis $\alpha \in \{X, Y, Z\}$, and $m$ is the number of encoding layers. This method allows encoding of large feature sets from network traffic into fewer qubits, reducing hardware requirements while retaining feature richness. Such encoding is particularly useful for large-scale IDS datasets.
\end{enumerate}

Table~\ref{tab:qfe_ids} gives an overview of feature encoding techniques for an efficient quantum IDS.

\begin{table*}[ht]
    \centering
    \caption{Summary of Quantum Feature Encoding Methods for Intrusion Detection Systems. Here, \(d\) denotes the number of classical input features extracted from network traffic data. The qubit requirements and complexities scale with \(d\) or \(\log d\) depending on the encoding scheme.}
    \label{tab:qfe_ids}
    \scriptsize
    \setlength{\tabcolsep}{4pt} 
    \renewcommand{\arraystretch}{1.1} 
    \begin{tabularx}{\textwidth}{lcXXX}
    \toprule
    \textbf{Encoding Method} & \textbf{Qubits Required} & \textbf{Suitable IDS Features} & \textbf{Pros} & \textbf{Challenges} \\
    \midrule
    Angle Encoding & \(O(d)\) & Continuous-valued metrics & Simple, hardware-friendly & Needs many qubits for high \(d\) \\
    \midrule
    Amplitude Encoding & \(O(\log d)\) & High-dimensional flow-based features & Compact representation & Complex state preparation \\
    \midrule
    Basis Encoding & \(O(d)\) & Binary/categorical attributes & Natural for discrete data & Not ideal for continuous features \\
    \midrule
    Phase Encoding & \(O(d)\) & Cyclic/temporal traffic patterns & Captures phase-based correlations & Requires entanglement for amplitude-based info \\
    \midrule
    Dense Angle Encoding & \(O(\log d) - O(d)\) & Mixed large feature sets & Encodes many features with fewer qubits & Circuit depth increases \\
    \bottomrule
    \end{tabularx}
\end{table*}

    \subsection{QML in Intrusion Detection}

    With growing interconnectedness between devices~\cite{Market}, the growth in data has been exponential. In the modern world, we rely on machine learning to help systems make automatic and smart decisions. Machine learning algorithms have made the process of decision-making very fast. One of the most common examples is detecting spam in emails. It could take days for a human to filter spam from a collection of emails by manually checking them whereas a ML algorithm can filter these spam in seconds. But as the amount and complexity of data is constantly growing, classical ML sometimes struggles.

    This is where \textbf{QML} comes in. It is reasonable to ask: \textit{"Why do we need something new like QML if classical machine learning already works?"}

    To address this question, it is important to understand the limitations of classical ML algorithms. Classical ML models work by processing combinations of input at a time which makes the model learn patterns point by point~\cite{faouzi2023classic}. This becomes computationally heavy if the data is huge and patterns are complex.

    Quantum computers process information differently, following the two main principles: superposition and entanglement. Superposition lets a quantum computer represent many combinations of inputs simultaneously. Imagine each data point is a box containing its features inside the box. Classical ML will look inside one box at a time whereas superposition enables QML to look inside multiple boxes simultaneously.

    In classical ML features are treated independently unless the model learns to link them, which takes a huge amount of data and time. In QML, the features are interdependent due to entanglement. The qubits are linked together, so their states become interdependent. If the state of one qubit changes, the states of the linked qubits will also change, no matter where they are. This helps in capturing correlations between features. This is especially helpful in IDS, where the interaction between network features represents an anomaly.

    These principles led to the integration of quantum algorithms into traditional machine learning workflows \cite{zeguendry2023quantum}. QML models such as quantum support vector machines (QSVMs) \cite{akrom2024quantum}, variational quantum circuits (VQCs) \cite{chen2020variational}, and quantum kernel methods \cite{schuld2021quantum}, can provide exponential or polynomial speed-ups in processing data. In the field of cyber security, QML has been investigated for anomaly detection, traffic pattern analysis and malware detection, where enhanced feature extraction and generalization capabilities are critical \cite{hdaib2024quantum, kukliansky2024network, wang2025network}.

    As technologies in quantum computing evolve, researchers are exploring the integration of QML into IDS. Most QML-based IDS fall into two main categories - VQC-based and quantum kernel-based~\cite{bhowmik2024quantum}. VQC consists of a PQC that uses a sequence of quantum gates controlled by tunable classical parameters.  The parameters of PQC are optimized in classical computers to reduce the cost function. On the other hand, kernel-based methods make use of quantum computers to encode the data into high-dimensional Hilbert spaces. The similarities between data points are measured using inner products of quantum states. The quantum kernel captures how similar the two inputs are in this space. A decision boundary is defined by suitable quantum measurement, allowing complex patterns (like anomalies) to be separated linearly in the transformed space. 

    \begin{figure*}[ht]
            \centering
            \includegraphics[width=1\linewidth]{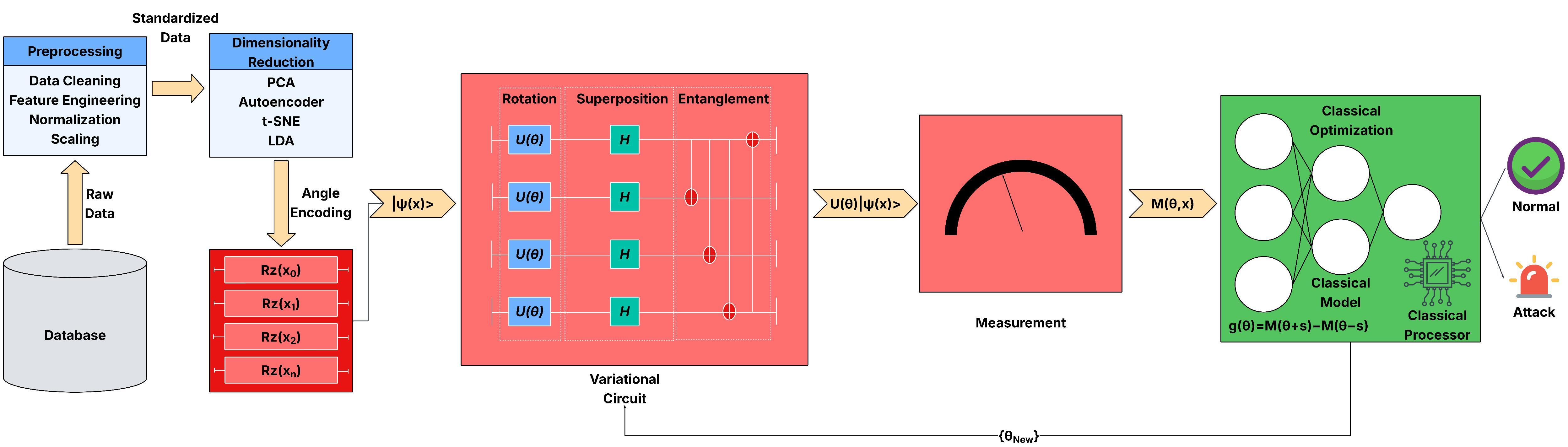}
            \caption{Overview of the QML workflow for intrusion detection on network data. The raw network data is first preprocessed and dimensionally reduced, followed by encoding into a quantum circuit. A classical optimizer is then applied to train the quantum model, and finally, the processed data is classified to detect intrusions.}
            \label{fig:qml}
    \end{figure*}
    
    The quantum algorithms are classified into supervised, unsupervised and reinforcement learning.

    \subsubsection{Supervised Quantum Algorithm} In supervised QML, labeled encoded classical data is passed through a quantum circuit that outputs a probability distribution over different classes. The outputs are measured after each iteration to get the probability distribution of different classes. Some of the most common supervised quantum algorithms are:

    \begin{itemize}
        \item Quantum Support Vector Machine (QSVM): QSVM is the counterpart of classical SVM. SVM is based on a hyperplane that separates data points of different classes with maximum margin. QSVM is based on a similar approach, but it performs the computation in a high-dimensional Hilbert space using quantum states and quantum kernels. Quantum computers can evaluate the inner products faster, leading to faster evaluation of kernel matrix~\cite{rebentrost2014quantum}. The QSVM can encode classical data into exponentially large Hilbert spaces~\cite{schuld2019quantum}, enabling the separation of classes that are not linearly separable in classical classes. The attacker will make sure that the anomalous data looks like normal data to not get caught, which makes classical SVM separate the normal/anomalous class challenging. However, in QSVM, the encoded data in high-dimensional space makes a clear boundary between normal/anomalous data facilitating more accurate and efficient classification.

        Consider a dataset of network traffic feature vectors 
\[
\{(\mathbf{x}_i, y_i)\}_{i=1}^M,
\]
where each \(\mathbf{x}_i \in \mathbb{R}^d\) represents the \(d\)-dimensional feature vector of a network flow, and 
\(y_i \in \{+1, -1\}\) denotes the class label, with +1 for normal traffic and -1 for anomalous traffic. 

The QSVM employs a parameterized quantum circuit \(U_{\Phi}(\mathbf{x})\) to construct a quantum feature map 
\[
|\phi(\mathbf{x})\rangle = U_{\Phi}(\mathbf{x}) |0\rangle^{\otimes n},
\]
which embeds classical data into an exponentially large Hilbert space using \(n\) qubits. 

The similarity between two inputs \(\mathbf{x}\) and \(\mathbf{x}'\) is computed via a quantum kernel defined as the squared inner product of their encoded quantum states: 
\[
K(\mathbf{x}, \mathbf{x}') = \left| \langle \phi(\mathbf{x}) | \phi(\mathbf{x}') \rangle \right|^2
\]

Using this kernel, the QSVM solves a classical quadratic optimization problem to find the optimal Lagrange multipliers \(\alpha_i\) that maximize the margin between classes, subject to the constraints 
\[
0 \leq \alpha_i \leq C, \quad \sum_{i=1}^M \alpha_i y_i = 0,
\]
where \(C\) is a regularization parameter controlling the trade-off between margin maximization and classification error. 

The resulting decision function for anomaly detection predicts the class of a new input \(\mathbf{x}\) by evaluating
\[
f(\mathbf{x}) = \text{sign} \left( \sum_{i=1}^M \alpha_i y_i K(\mathbf{x}_i, \mathbf{x}) + b \right),
\]
where \(b\) is the bias term learned during training.

    \item Quantum Neural Network (QNN): A QNN is a variational circuit with learnable parameters, which are to be optimized while training~\cite{qiskit_nn_tutorial}. The gates in variational circuits are controlled by tunable parameters. These tunable parameters are adjusted during training like weights in classical ML. Fig.~\ref{fig:qml} shows an overview of QML in intrusion detection systems.Designing quantum circuits for intrusion detection include several key steps:
    \begin{enumerate}
        \item Feature Encoding: Network traffic features are first encoded using encoding techniques (as mentioned in Section~\ref{feature_maps}) so that they can be projected into high-dimensional Hilbert space as quantum states.
        \item Parameterized Quantum Layers: The quantum circuit comprises layers of parameterized gates (e.g., rotations around X, Y, or Z axes) whose parameters are trainable. These layers act similarly to hidden layers in classical neural networks, allowing the QNN to learn complex nonlinear decision boundaries.
        \item Entanglement: Entangling gates (such as CNOT) are strategically placed between qubits to capture correlations among different network features. This entanglement enhances the expressive power of the network, helping it model intricate relationships typical in network traffic.
        \item Measurement and Output: After the parameterized transformations, measurements on one or more qubits produce probabilistic outputs. These outputs correspond to class probabilities or decision scores indicating whether a network flow is normal or anomalous.
        \item Training: Classical optimization algorithms update the quantum gate parameters based on a loss function computed from measurement outcomes, typically using gradient-based methods adapted for quantum circuits.
    \end{enumerate}

    By embedding network traffic data into quantum states and exploiting quantum phenomena like superposition and entanglement, QNNs can explore complex feature spaces more efficiently than classical counterparts. This capacity enables QNNs to detect subtle anomalies and evolving attack patterns that traditional methods may miss.

        \item Quantum k-Nearest Neighbor (QkNN): The classical kNN algorithm is based on a distance metric (e.g. Euclidean distance). This algorithm identifies the k closest points (neighbors) to a query data point based on a distance metric and uses their label to predict the label of the query data. However, when the data size and dimensionality increase, the classical kNN becomes computationally very expensive due to the need to calculate distances between all the data and a query data point. In QkNN, all training data can be encoded into a superposition of quantum states. Unlike in classical kNN where looping is required on $n$ states one by one, the quantum algorithm can process all of them at once in a single operation using quantum parallelism. The distance in QkNN can be measured using fidelity~\cite{basheer2020quantum}, which is related to inner products and is given by:
        \[
        F(\psi_q, \psi_i) = \left| \langle \psi_q | \psi_i \rangle \right|^2.
        \]
        The fidelity is computed using quantum circuits such as swap tests. Mathematically, fidelity quantifies the overlap between two quantum states, with values ranging from 0 (completely different) to 1 (identical). This measure is crucial for distinguishing subtle differences in network behavior, especially when anomalous traffic closely resembles normal traffic patterns to avoid detection. QkNN can alos be implemented using various distance metrics (such as Hamming and Euclidean distance)~\cite{li2022quantum, zardini2024quantum}. In IDS, the query network traffic can be compared with the labeled traffic data using fidelity or various other distance metrics to classify it as normal or anomalous. QkNN can help detect anomalous network data faster as compared to classical kNN due to quantum parallelism.

        \item Hybrid Quantum Classical Neural Network (HQCNN): The HQCNN is a combination of quantum and classical neural networks~\cite{arthur2022hybrid, ling2024image}. In HQCNN, the feature dimension of classical data is reduced by passing through a classical neural network. The output of the first classical neural network layer acts as input to the quantum circuit. The output is encoded as quantum states using methods like angle or amplitude encoding so that quantum algorithms can operate on it. The quantum operation happens in the second layer. The output of the quantum layer is measured and fed back to the classical neural network layer to get the desired output. This method can efficiently detect anomalies in network traffic where network features have interdependent patterns. Quantum entanglement can help capture dependencies between different types of network behaviors.
        
    \end{itemize}

    \subsubsection{Unsupervised Quantum Algorithm} Unsupervised QML makes use of unlabeled data. In anomaly detection, it's very difficult to always get labeled data. This QML learns from the data without prior knowledge. Unsupervised methods are more used in intrusion detection due to the scarcity of labeled data. Some of the most common unsupervised quantum algorithms are:
    \begin{itemize}
        \item Quantum Autoencoder (QAE): Classical autoencoders encode classical data into a lower dimension and learn to reconstruct the original classical data from this lower dimension~\cite{li2023comprehensive}. The training is done between the input and output of this algorithm to reduce the reconstruction loss. Similarly, QAE encodes the quantum data stored on n qubits into a smaller quantum register of $m<n$ qubits~\cite{romero2017quantum}(see Fig.~\ref{fig:qae}). In quantum computing, the qubits cannot be simply erased/discarded due to quantum operations being unitary (i.e., reversible). Suppose an initial quantum state living in Hilbert space is given by:
        \[
        H = H_{C} \otimes H_{T},
        \]
        where, $H_{C}$ space for compressed (coded) qubits (useful information) and $H_{T}$ space for trash qubits (irrelevant data).

        The objective is to find a unitary transformation $U$ such that:
        \[
        U|\psi \rangle = |\phi \rangle_{C} \otimes |0\rangle_{T},
        \]

        where \( |\phi \rangle_{C} \) represents the compressed (coded) state and \( |0\rangle_{T} \) denotes the trash register initialized to a known basis state. After compression, the goal is to reconstruct the original state from the compressed state. To achieve this, inverse transformation \( U^\dagger \) is applied to the compressed state with trash state re-initialized to $|0\rangle$:
        
        \[ U^\dagger(|\phi\rangle_{C} \otimes |0\rangle_{T}) = |\psi\rangle. \]

        After applying $U$ and $U^\dagger$, the closeness between the final output and the original input $|\psi\rangle$ is measured. The loss function is often fidelity and is given by:
        \[
        \mathcal{L} = 1 - |\langle \psi | U^\dagger U|\psi\rangle|^{2
        }.
        \]

        Minimizing the loss ensures that the encoder-decoder pair accurately reconstructs the input. The QAE training updates the parameters in the PQC based on the comparison between the input and output states. QAE can be very helpful in IDS. As it is an unsupervised learning method, it does not require labeled data. The QAE can be trained to reconstruct the normal behavior of network data so that whenever any anomaly is detected, the increase in reconstruction loss can help in detecting such anomalies in the network. QAE can leverage the exponentially large Hilbert space of quantum states to efficiently compress high dimensional and complex network traffic data. 
        \begin{figure}[t]
            \centering
            \includegraphics[width=1\linewidth]{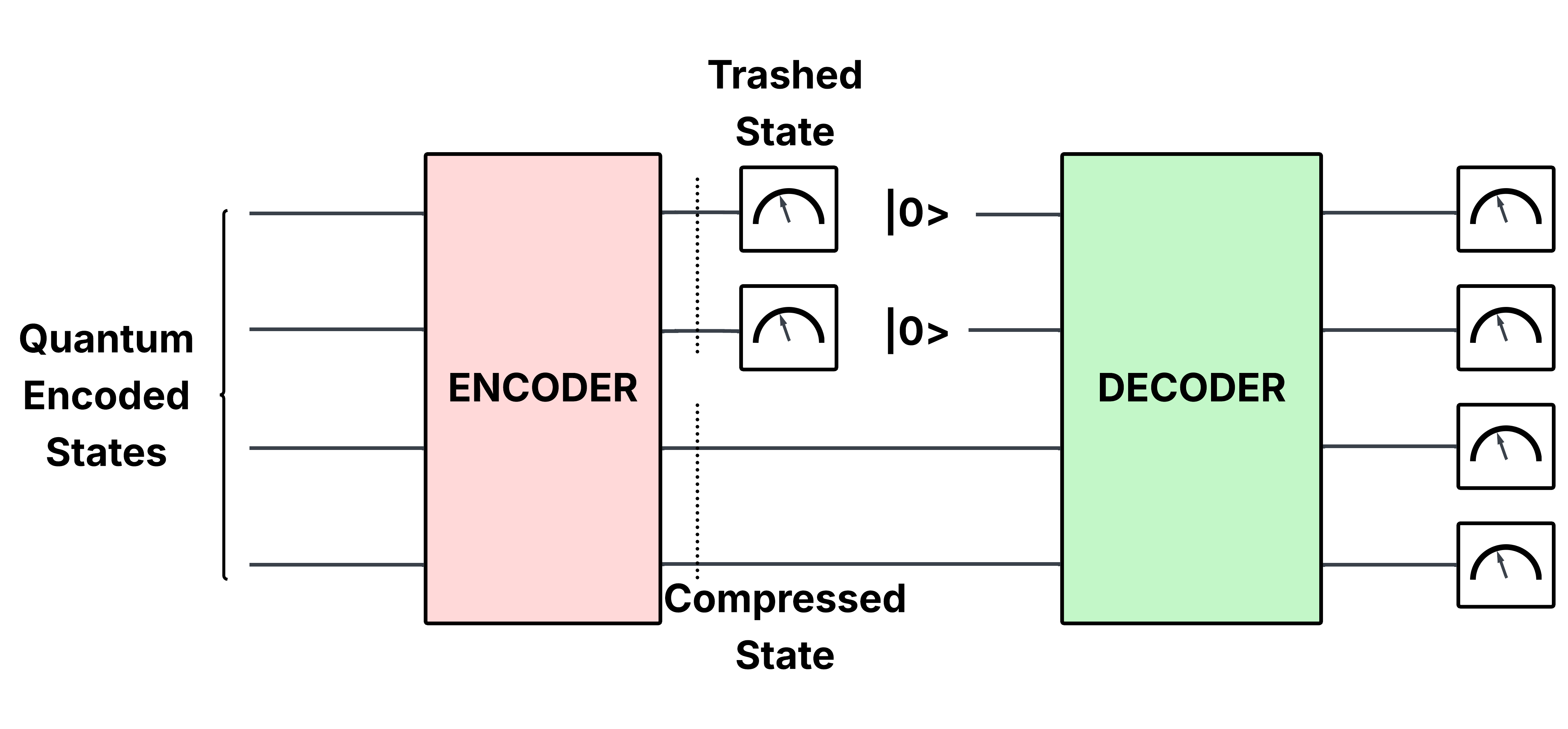}
            \caption{Quantum Autoencoder}
            \label{fig:qae}
        \end{figure}
        
        \item Quantum Clustering: Quantum clustering algorithms are designed to group data into meaningful clusters by exploiting quantum mechanical phenomena. Unlike classical clustering methods (such as k-means or DBSCAN), quantum clustering can map data into a high-dimensional Hilbert space using quantum embeddings, that can capture complex data structures that may be inaccessible by classical techniques. Quantum clustering leverages quantum states, quantum kernels, or quantum-inspired optimization techniques to perform clustering. Firstly, the classical data is mapped to quantum states into quantum Hilbert space. Then, various distance/similarity measures (such as fidelity/Hilbert-Schmidt inner product/quantum kernel functions) are used to measure the similarity between quantum states. If the distance between the data points is less or the similarity is more, they are assigned to the same clusters.

        In~\cite{10.1145/1273496.1273497}, A\"{\i}meur et al. introduced thee Grover-based quantum subroutines to accelerate classical clustering tasks. These subroutines include quant-find-max (used to find the pair of points in the dataset with maximum distance), quant-find-c-smallest-values (which identifies the $c$ nearest neighbors of a given point), and quant-cluster-median (a novel quantum routine that determines the median of a point set by computing the total distance from each point to all others and then applying Grover’s minimum search to identify the point with the smallest overall distance). Using these primitives, authors developed three clustering algorithms: \textit{quantum divisive clustering, $k$-medians and construction of a $c$-neighborhood graph}.

        Divisive clustering starts by considering all the data points in the same cluster. The second step is to split this cluster with all the data points into two sub-clusters. This is done by finding two data points that are farthest apart within the cluster and are called seeds. All other data points are attached to their nearest seeds. This method is applied recursively until the clusters with similar data points are created.

        The $k$-median (also called $k-$medoids), is similar to $k$-means clustering algorithm. Instead of calculating the mean of the data points to define its center as in $k$-means clustering, it uses the real median value of the data point that minimizes the sum of distances to all other points in the cluster. It is an iterative algorithm and stops when the center of the clusters has stabilized.

        The $c$-neighborhood graph method starts by considering all the data points as vertices of a complete graph and edges between these vertices are weighted by the distance between these two data points. A $c$-neighborhood graph is obtained by discarding all the edges except the edges linking to its $c$ closest neighbors.

        All these algorithms are based on Grover's algorithm~\cite{grover1997quantum} which attains quadratic speedup compared to classical methods. Suppose there are $N$ unsorted items, classically it would take $N/2$ steps to search items on average and $N$ steps in the worst case. Grove's algorithm can find the solution in $\sqrt{N}$ steps by leveraging quantum parallelism. The ability of quantum computing to handle high-dimensional data and quadratic speedup achieved in Grover's algorithm shows its potential in network attack detection, particularly suitable for real-time and adaptive security applications.
        
        \item Quantum Ensemble Methods: They extend classical ensemble methods, such as bagging, boosting, and stacking, into the quantum domain~\cite{macaluso2020quantum}. In classical learning, many weak learners are used to develop a strong learner that outperforms a single model. Similarly, in quantum computing, quantum ensemble techniques leverage the quantum probabilistic nature and parallelism of quantum computing to produce diverse models and aggregate them to develop a strong model with enhanced performance~\cite{schuld2018quantum}. Quantum ensemble methods can capture richer patterns in network traffic using fewer resources compared to the classical model. Classical ensemble methods often struggle with high-dimensional and complex network flow data whereas quantum models can process high-dimensional and complex data in exponentially smaller space which makes quantum ensemble methods more suitable for IDS. It can help reduce false positives and accurately detect anomalies in the network.
        
        \item Quantum Boltzmann Machine (QBM): It is a quantum-enhanced version of the classical Boltzmann machine that leverages the principles of quantum mechanics. The classical energy function is replaced with a quantum Hamiltonian, and the system's state is described by a thermal quantum density matrix in a QBM~\cite{amin2018quantum}. This allows a QBM to model high dimensional probability distribution more efficiently as compared to the classical Boltzmann machine. The patterns in network traffic are very intricate. These patterns can make classical Boltzmann machine hard to learn complex patterns. The QBM can help reduce false positives and detect real-time anomalies due to high-dimensional feature learning which can be very advantageous in IDS.

        \item Quantum Generative Adversarial Network (qGAN): The qGAN combines the power of GANs learning paradigm and quantum computation~\cite{qiskit_nn_tutorial}. The qGANs use a quantum generator $G_{\Theta}$ to generate synthetic data distributions that closely resemble original data and discriminator $D_{\phi}$ to distinguish between original and generated data distributions. Both generator and discriminator are trained together in a min-max game where $G_{\Theta}$ goal is to fool the discriminator by generating so realistic data that $D_{\phi}$ cannot tell it's fake and $D_{\phi}$'s goal is to classify the data as real or fake. Lloyd et al. in~\cite{lloyd2018quantum} first proposed the concept of qGAN, showing that quantum generators could be more expressive than classical ones, especially when operating in high-dimensional Hilbert spaces. This adversarial learning can be particularly advantageous for intrusion detection, where labeled data is scarce and anomalies are rare, making it suitable for unsupervised or semi-supervised learning. By learning the distribution of normal traffic, qGANs can flag any deviations from normal behavior as a potential threat.
        
        \item Quantum PCA (QPCA): QPCA is a quantum analogue of classical PCA, which is designed to reduce the dimensionality of quantum data while preserving the most significant features. It was first introduced by Lloyd et al.~\cite{lloyd2014quantum}, who proposed an efficient quantum algorithm that estimates the principal components of a density matrix using quantum phase estimation and controlled unitary operations. The classical PCA scales poorly with high-dimensional data as it requires eigen-decomposition of the covariance matrix. QPCA can leverage the exponential speedup potential of quantum computing to analyze large datasets. QPCA can be more efficient in processing network traffic data due to the massive feature space. It can preprocess data by reducing irrelevant or redundant features, thereby enhancing efficiency.
    \end{itemize}
    \subsubsection{Quantum Reinforcement Learning (QRL)} As in classical reinforcement learning, in QRL, the agent learns to make decisions by interacting with an environment to maximize cumulative reward. Like traditional reinforcement learning (RL), a QRL is built with three components:  a policy, a reward function, and a model of the environment~\cite{dong2008quantum}. However, QRL differs from RL in how it represents states and actions using quantum states in high-dimensional Hilbert space. Each action or state can exist in a superposition of multiple eigen states allowing exploration of many possibilities simultaneously. Action in QRL is based on quantum measurement and values across all states can be updated simultaneously utilizing quantum parallelism. The learning process involves updating the probability amplitudes of actions based on received rewards. In IDS, where network traffic data is complex and high-dimensional, traditional RL methods may struggle with scalability and precision. QRL utilizing quantum parallelism can be highly scalable and efficient in learning high-dimensional features and intricate patterns of network traffic data.
    
    \subsection{Case Studies of QML in IDS}

    \begin{table*}[ht]
        \centering
        \caption{Summary of Case Studies in QML for IDS}
        \label{tab:qml_ids}
        \begin{tabular}{p{3.5cm}p{2cm}p{2.8cm}p{2.5cm}p{2.5cm}}
        \toprule
        \textbf{References} & \textbf{Dataset(s)} & \textbf{Quantum Model}  & \textbf{Platform/Tools} & \textbf{Performance Metrics} \\
        \midrule
        Abreu et al.~\cite{abreu2025quantumnetsec} & UNSW-NB15, CIC-IDS17, CICIoT23, TON IoT & VQC, QKM, QCNN, QSVM & Qiskit & F1 score   \\
        \midrule
        Kumar et al.~\cite{kumar2025quids} & EDGE-IIOTSET, ACI IoT & QSVC & Qiskit &  Recall, Precision, F1 score \\
        \midrule
        Hdaib et al.~\cite{hdaib2024quantum} & KDD99, IoT-23, CIC-IoT-23 & QAE with QSVM, QAE with Quantum Random Forest, QAE with QkNN & Qiskit, PennyLane PyTorch & Accuracy, Precision, Recall, F1 score \\
        \midrule
        Rahman et al.~\cite{rahman2023quantum} & NSL-KDD & qGAN  & Qiskit, PyTorch &  Cumulative Distribution Function (CDF) \\
        \bottomrule

        \end{tabular}
    \end{table*}

    QML can be an efficient solution for detecting network anomalies. This section discusses four different works on network anomaly detection using QML.

    In~\cite{hdaib2024quantum}, Hdaib et al. proposed three different quantum autoencoder (QAE) frameworks to detect anomalies in network traffic. The three proposed frameworks are formed by integrating QAE with a quantum one-class support vector machine, a quantum random forest and a quantum $k$-nearest neighbor, and are tested on three different benchmark datasets comprising network flows. A QAE can be trained on "normal" data samples to reconstruct the original data from the compressed state. The primary goal is to minimize the loss of reconstruction. The high reconstruction loss shows an anomaly in the network flow. The QAE alone achieved an accuracy of 75\% and an F1 score of 77\%. However, when integrated with other QML methods, it significantly outperformed the baseline, demonstrating high detection capabilities and robustness. Among the three proposed frameworks, QAE when combined with QkNN achieved the best performance, with an accuracy of 97.79\%, and an F1 score of 98.26\%. The authors concluded that quantum deep learning when integrated with parameterized quantum circuits can significantly enhance anomaly detection performance.

    In~\cite{kumar2025quids}, Kumar et al. developed an IDS for IoT networks based on a quantum support vector classifier (QSVC) namely QuIDS. Two publicly available IoT network traffic datasets were used. The architecture of the proposed module included three main modules: Flow Reconstruction, Feature Extraction, and QSVC Classification. The input to the flow reconstruction module was a set of packets $P = \{P_{1},P_{2},P_{3},...,P_{n}\}$ and the output was the set of flows $F = \{F_{1},F_{2},F_{3},...,F_{m}\}$ were $n>m$. The set of flows, F, is given as input to the feature extraction module which processes these flows and outputs flow-level features vector given by $X = \{\vec{x_{1}},\vec{x_{2}},\vec{x_{3}},...,\vec{x_{8}}\}$. This module extracts eight flow-level generic traffic features such as Flow Direction Ratio (FDiR), Flow Payload Ratio (FPR), Flow Duration Ratio (FDuR), Flow Packet Rate Ratio (FPRR), Flow Inter-arrival Time Ratio (FITR), Flow Packet Length Ratio (FPLR), Flow Length Rate Ratio (FLRR), and Flow TLS Record Length Ratio (FTRLR) that is to be utilized for attack classification in IoT network traffic. In the QSVC classification module, features are reduced to optimal qubits using PCA. The features are mapped to quantum states using three different feature mapping techniques (i.e., ZFeature, ZZFeature and PauliFeature maps). After encoding features to quantum states, the QSVC was trained using the COBYLA optimizer. The results were compared with five machine learning, three deep learning and two recent state-of-the-art methods. The results showed that QuIDS using the ZZFeature map outperformed all the other compared models in terms of precision, recall and F1-score. 

    Rahman et al. in~\cite{rahman2023quantum} implemented a quantum generative adversarial network (qGAN) using PyTorch in conjunction with Qiskit. In the proposed architecture, the generator was modeled as a quantum circuit and the discriminator remained classical. Binary cross-entropy was used as a loss function which is given by:
    \[
L(\theta) = \sum_j p_j(\theta) \left[ y_j \log(x_j) + (1 - y_j) \log(1 - x_j) \right]
\]
    where $x_j$ and $y_j$ refer to instances and to the corresponding label.
    
    The authors plotted the cumulative distribution function of the original samples, generated samples and the absolute difference between them. This implementation revealed that quantum algorithms are capable of capturing nuanced relationships in network traffic data that classical methods often miss.

    In~\cite{abreu2025quantumnetsec}, Abreu et al. proposed QuantumNetSec, a QML-based network security framework specifically tailored for Noisy Intermediate-Scale Quantum (NISQ) devices. The QuantumNetSec algorithm begins with monitoring real-time network traffic and metadata collection. The raw data is preprocessed using classical methods and mapped to quantum states using a feature map. It is followed by transpilation and circuit optimization techniques adapted to the NISQ backend. The proposed algorithm was tested using four different QML models: variational quantum circuits (VQC), quantum k-means (QKM), quantum convolutional neural networks (QCNN), and quantum support vector machines (QSVM). These quantum models were optimized using classical optimizers. In the proposed method four feature maps (RawFeatureVector, PauliFeatureMap, ZFeatureMap and ZZFeatureMap), four Ansatz (RealAmplitudes, EfficientSU2, ExcitationPreserving and TwoLocal), and four different optimizers (ADAM, COBYLA, SPSA and gradient descent) were employed. A key innovation in this method is the personalized optimization technique which tests all the combinations of transpilation methods, such as init, layout, translation, routing, optimization and scheduling, to reduce the noise in the quantum hardware. For evaluation, seven different backends from the IBM quantum platform were used: Cairo, Kyoto, Brisbane, Osaka, Sherbrooke, Torino, and Quebec. The results showed that the personalized optimization technique increased the performance of all four QML models, achieving a higher F1 score, and QCNN achieved the highest F1 score in both binary and multiclass classification.

    Table~\ref{tab:qml_ids} provides a summary of case studies in QML methods for intrusion detection systems.


    \section{Transitioning from QML to QFL}

    As quantum computing advances, QML has emerged as a great solution that can leverage quantum advantages to gain exponential speedups and achieve efficient learning on high-dimensional data. Various QML algorithms have shown empirical benefits over their classical counterparts for specific tasks. However, the majority of QML applications consider centralized settings which limit their applicability and scalability in privacy-sensitive domains such as healthcare, finance, and network security.

    On the other hand, FL frameworks ensure the privacy and security of the collected data by training the model on the device itself, eliminating the need to transfer the actual raw data for centralized training. These frameworks and their applications have been widely studied in classical machine learning, but it is now increasingly becoming relevant in the quantum domain. The integration of QC with FL - known as \textit{Quantum Federated Learning (QFL)} seeks to harness the exponential speedup of quantum computing while maintaining the decentralized and privacy-preserving nature of FL. In terms of network security, sharing raw traffic across organizations or devices may violate user privacy or regulatory constraints. In such a scenario, QFL can help to detect anomalies in network traffic faster and ensure secure quantum communication between local and global devices.

    QFL follows similar methods as classical FL:
    \begin{enumerate}
        \item Each quantum client $n$ will have datasets $D_{n}$ for training local model.
        \item The clients train local models $\theta_{n}$ to minimize a loss function, $\theta_n = \arg \min_{\theta} \mathcal{L}(\theta, D_n) $.
        \item The local models are aggregated by the server to produce a global model, $\theta_{global} = Aggregate (\theta_1, \theta_2, ..., \theta_n)$.
        \item The local models are sent back to each client for further training.
    \end{enumerate}

    Each client trains a PQC with trainable classical parameters (such as gate rotation angles). Since trainable parameters are classical, they can be aggregated by using classical FL model aggregation techniques mentioned in Section~\ref{sec:model_aggregation}. Although, the underlying model uses quantum circuits, the trainable parameters—such as gate rotation angles in PQCs—are, in most current implementations, classical values.

    Table~\ref{tab:qfl_platforms} provides a concise summary of prominent platforms supporting the development of QFL solutions. 

\begin{figure*}[ht]
        \centering
        \includegraphics[width=1.00\linewidth]{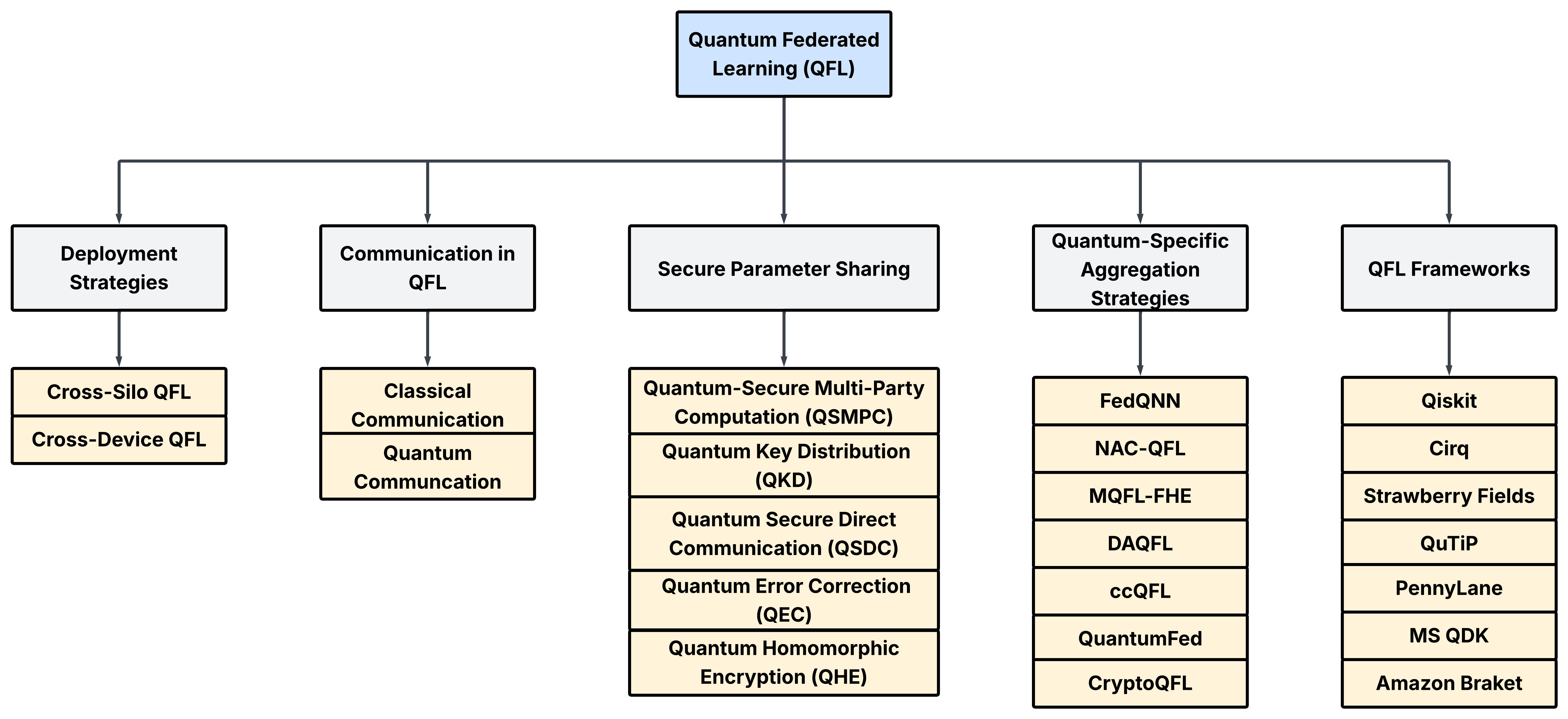}
        \caption{Taxonomy of QFL for intrusion detection, illustrating the hierarchical organization of QFL aspects including deployment strategies, communication mechanisms, secure parameter sharing, quantum-specific aggregation methods, and existing QFL frameworks.}
        \label{fig:qfl}
    \end{figure*}

    Fig.~\ref{fig:qfl} presents a taxonomy of QFL. The taxonomy organizes QFL approaches into key dimensions such as deployment strategies, communication, aggregation, privacy and QFL frameworks. This structured overview highlights the major design directions and helps position existing research within the broader QFL landscape.

\begin{table*}[ht]
\centering
\small
\caption{Comparison of Quantum Computing Platforms for QML and QFL}

\begin{tabular}{@{}l l l l p{2cm} p{6.5cm}@{}}
\toprule
\textbf{Platform} & \textbf{Developer} & \textbf{Language} & \textbf{QML} & \textbf{QFL} & \textbf{Key Features / Remarks} \\
\midrule
\textbf{Qiskit}~\cite{cross2018ibm} & IBM & Python & Yes & Yes & Terra, Aer, Aqua; supports hybrid QML models and integration with FL via PySyft or Flower. \\
\midrule
\textbf{Cirq}~\cite{omole2020cirq} & Google & Python & Yes & Possible & Low-level gate model; integrated with TensorFlow Quantum; flexible for hybrid ML. \\
\midrule
\textbf{Strawberry Fields}~\cite{killoran2019strawberry} & Xanadu & Python & Yes & Experimental (via PennyLane) & Photonic (CV) quantum computing; used with PennyLane for hybrid QML. \\
\midrule
\textbf{QuTiP}~\cite{johansson2012qutip} & Open-source & Python & Limited & No & Simulates open quantum systems; not designed for ML or FL. \\
\midrule
\textbf{PennyLane}~\cite{bergholm2018pennylane} & Xanadu & Python & Strong & Yes (via Flower, PySyft) & Hybrid QML; supports Braket, Qiskit, Cirq; used in QFL research with Flower/PySyft. \\
\midrule
\textbf{TF Quantum}~\cite{broughton2020tensorflow} & Google & Python & Yes & Possible (via TensorFlow Federated) & TensorFlow + Cirq for hybrid models; extendable to QFL via TensorFlow Federated. \\
\midrule
\textbf{MS QDK}~\cite{QDK} & Microsoft & Q\# & Basic & No & Designed for algorithm prototyping in Q\#; lacks ML/FL support. \\
\midrule
\textbf{Amazon Braket}~\cite{braket} & AWS & Python & Yes & Yes & Access to IonQ, OQC, Rigetti; PennyLane integration; QFL possible via custom FL orchestration. \\
\bottomrule
\end{tabular}
\label{tab:qfl_platforms}
\end{table*}

     \subsection{QFL Deployment}

     \subsubsection{QFL for Cross-Silo Network Security}

    Cross-Silo QFL lets various organizations (e.g., banks, government agencies or ISPs) collaboratively detect intrusions or attacks without sharing raw network traffic. Each organization has substantial infrastructure (HPC clusters, quantum simulators, or priority access to cloud QPUs). Each silo collects network traffic from its own perimeter. These classical network traffic data are mapped to quantum states using feature mapping techniques, and a quantum model is trained locally across each organization. Only quantum circuit parameters or model weights are securely aggregated at a central coordinator. Techniques such as QKD~\cite{scarani2009security} and quantum secure multi-party computation~\cite{crepeau2002secure} can be incorporated to ensure the security of the shared-parameters while communicating with the server.

    \subsubsection{QFL for Cross-Device Network Security}

    As the number of connected devices continues to grow, edge nodes are generating and capturing massive volumes of data. The data at edge devices can be highly complex and non-IID. Most edge devices—such as IoT cameras, routers, and smart meters—work collaboratively to detect network anomalies at the local level. The edge nodes have various resource constraints, including limited power, computational ability, and network connectivity, which often makes them rely on cloud quantum access. In the cross-device setting, most existing QFL approaches remain at the simulation or prototype stage, with limited deployment in real-world IoT environments.

    \subsection{Communication in QFL}

    Communication in QFL can be purely classical (with quantum local training), fully quantum (with quantum state sharing), or hybrid. The choice depends on hardware constraints, desired security, and application domain. In the near term, classical communication secured with QKD~\cite{mao2018integrating, qi2016simultaneous} is most practical, whereas entanglement-assisted quantum communication represents the long-term vision.

    In QFL, the communication happens in two main ways:
    \begin{enumerate}
        \item Client $\rightarrow$ Server: Sending quantum, classical or hybrid updates.

        \item Server $\rightarrow$ Clients: Broadcasting the global model (quantum circuit parameters).
    \end{enumerate}
    To explore these modalities in more depth, we distinguish between 
\textit{classical communication} and \textit{quantum communication} in QFL~\cite{nguyen2025quantum}.
\begin{itemize}
    \item Classical Communication: In classical communication settings, clients train quantum models locally 
but exchange only \emph{classical information} with the server. This typically 
includes circuit parameters (e.g., rotation angles), gradient information, 
or measurement outcomes. The server then aggregates these classical updates 
using protocols such as FedAvg, before redistributing the global parameters 
back to clients. Security can be enhanced through QKD, which ensures that 
updates are transmitted over classically accessible channels with 
quantum-secured encryption.
    \item Quantum Communication: 
    Quantum communication enables clients to transmit \emph{quantum states} or 
quantum-encoded updates directly to the server. This approach leverages 
resources such as entanglement distribution, quantum teleportation, or 
quantum homomorphic encryption. Unlike classical communication, it avoids 
the potential leakage of private information contained in intermediate 
classical representations. However, practical implementation faces challenges 
including noise, decoherence, and the current limitations of quantum network 
infrastructure.
\end{itemize}

    \subsection{Secure Parameter Sharing in QFL}

    An efficient and secure communication is one of the major challenges of FL. Classical encryption techniques provide a certain level of protection, but they may fail with increasingly sophisticated attacks. Quantum-secure multi-party computation (QSMPC)~\cite{crepeau2002secure} approaches in QFL play a foundational role by enabling multiple parties to jointly compute model updates without revealing their private data. Building upon this, protocols such as Quantum Key Distribution (QKD)~\cite{scarani2009security} and Quantum Secure Direct Communication (QSDC)~\cite{zhang2017quantum} leverage quantum mechanics principles, like the irreversibility of measurement and photon polarization, to safeguard against eavesdropping by detecting interception through elevated quantum bit error rates. Additionally, quantum teleportation~\cite{bouwmeester1997experimental} allows the transfer of quantum states without physically transmitting the qubits themselves. Using pre-shared entanglement combined with classical communication, teleportation facilitates highly secure quantum parameter sharing, especially over long or noisy channels. Finally, quantum techniques such as quantum error correction (QEC)~\cite{terhal2015quantum} can be utilized to preserve the integrity and reliability of quantum data exchanged via teleportation, QKD, or QSDC. Various quantum techniques can be used to compress the model to reduce the communication overhead.

    Another technique for secure communication can be quantum homomorphic encryption (QHE), a cryptographic technique that allows quantum computations to be performed directly on encrypted quantum data without needing to decrypt it first. In QFL, where multiple parties collaboratively train quantum or quantum-classical hybrid models, QHE allows model updates and aggregations to be computed securely by a central server or aggregator without compromising the privacy of individual clients’ quantum data. This enhances trustworthiness and security in federated environments.

    \subsection{Quantum-Specific Aggregation Methods for QFL}

    \subsubsection{FedQNN} The FedQNN framework proposed by Innan et al. in~\cite{innan2024fedqnn} extends federated learning by integrating quantum neural networks (QNNs) to enable collaborative model training while maintaining local data privacy. In this approach, each client retains its private dataset and encodes it into quantum states processed through PQCs. The quantum model updates, represented as optimized quantum parameters, are communicated to a central server rather than the raw data. The server aggregates these updates to form a global quantum model, mathematically expressed as the average of client parameters, which enhances overall predictive performance and accuracy. Within the QNN, Hadamard gates create superposition, rotation gates introduce phase shifts and superpositions along X and Y axes, and the $U(\theta)$blocks iteratively optimize network parameters. Measurements on the final quantum state produce predictions, which are evaluated using a mean squared error (MSE) loss function. FedQNN also incorporates secure communication protocols and privacy-preserving aggregation mechanisms, allowing clients to contribute to the global model without exposing sensitive data. This integration of QNNs with federated learning leverages the computational advantages of quantum circuits while addressing the challenges of distributed and privacy-sensitive machine learning environments.

    \subsubsection{Noise Aware Clustered Quantum Federated Learning (NAC-QFL)} In~\cite{sahu2024nac}, Sahu et al. proposed NAC-QFL. The NAC-QFL system integrates a QML model with a parameterized quantum circuit into a federated learning framework across a set of quantum devices organized into clusters to optimize communication efficiency. Each cluster has a cluster head responsible for device selection, local model training, and coordination with a central aggregation server. Clustering uses a K-means algorithm with a distance metric combining inter-device communication distance and entanglement-assisted channel capacity, and cluster heads are selected based on both channel capacity and classical computational resources. The selection of devices within clusters is sensitive to noise, and the effective noise of each device is modeled considering qubit relaxation, dephasing, gate errors, measurement errors, and state preparation errors. The least noisy devices that satisfy circuit capacity and parallelization constraints are selected for training. During NAC-QFL training, the global model is distributed from the server to cluster heads, propagated to selected devices for local training, and aggregated back at the server, ensuring scalable, robust, and noise-resilient federated quantum learning.
    
\subsubsection{Multimodal Quantum Federated Learning
Framework with Fully Homomorphic Encryption (MQFL-FHE)} Dutta et al. in~\cite{dutta2024mqfl} proposed MQFL-FHE, which enables secure federated learning across heterogeneous data modalities, such as images, text, or biomedical sequences, while preserving client privacy. In this framework, each client preprocesses its local multimodal data and encodes it into quantum states using parameterized unitary operations. Local quantum model updates are computed via PQCs and subsequently encrypted using the CKKS fully homomorphic encryption (FHE) scheme, allowing the central server to aggregate encrypted model updates without accessing the raw data. Feature-level fusion is performed locally using attention-based mechanisms to capture correlations across modalities, and the aggregated global model is updated homomorphically by weighting client contributions according to their dataset size. This integration of quantum computation and FHE mitigates the performance degradation typically introduced by encryption while ensuring that sensitive data remains private during the federated learning process. The approach is scalable to multiple modalities and allows each client to contribute securely to a global model that captures multimodal relationships effectively. 

    \subsubsection{Dynamic Aggregation Quantum Federated Learning (DAQFL)} Qu et al. in~\cite{10869382} proposed DAQFL, which enables distributed clients, such as hospitals, to collaboratively train quantum neural networks (QNNs) while accounting for client-specific performance. Each client encodes classical data into quantum states via amplitude encoding and processes it through a VQC to obtain predictions. Local loss is computed, and circuit parameters are updated using parameter-shift rules. After training, clients upload gradients along with local accuracy to a central server. The server performs dynamic weighted aggregation, assigning higher weights to clients with better performance in the current round, computes the global gradient, and updates shared model parameters. Iterative local updates and aggregation continue until convergence, ensuring efficient, accuracy-aware, and privacy-preserving quantum federated learning.
    
    \subsubsection{Chained Continuous Quantum Federated Learning (ccQFL)} In~\cite{gurung2025chained}, Gurung et al. proposed ccQFL. In QFL, a central server is responsible for performing aggregation of parameters to produce a global model making it a single point of failure if the server is compromised. The constant communication between a central server and clients may not be feasible in the real world. ccQFL addresses this issue by performing serverless aggregation. Authors have introduced two variants of ccQFL. In the first variant, the devices are randomly grouped or selected based on some selection criteria. The devices train the models in parallel within the group. After the parallel training within the group is completed, the intra-group training is conducted sequentially without the need for an aggregation mechanism. In the second variant, $n$ devices are grouped in pairs. The model is first trained on the first device of each pair and passed on to the second device of the same pair for further training. The trained model from the second device is passed to the untrained second device in alternate pairs. Finally, the models of these devices are averaged and the process continues until the training is completed. This method ensures serverless aggregation of the model and also helps in dealing with heterogeneity and non-IID data.
    
    \subsubsection{QuantumFed} Xia et al. in~\cite{9685012} proposed QuantumFed framework that extends federated learning into the quantum domain, enabling collaborative training of quantum neural networks (QNNs) across distributed nodes while preserving local data privacy. In this framework, each node encodes its local dataset into quantum states and processes them through parameterized perceptron unitaries, which serve as the quantum analogue of classical model weights. Local updates are performed by computing update matrices via gradient-ascent-like procedures based on fidelity as the cost function, measuring the closeness between output and label quantum states. These update unitaries are applied iteratively over a defined interval length, and the resulting updates are communicated to a central server. The server performs global updates by aggregating the update unitaries from selected nodes, effectively maximizing the fidelity-based cost function across all participating nodes. The framework leverages the multiplicative identity property of unitary updates to ensure that the order of aggregation minimally impacts convergence, and the process can be mathematically interpreted as averaging the partial traces of local updates. QuantumFed thus allows each node to optimize its QNN locally while contributing securely to a global quantum model, combining the advantages of quantum computation with federated learning principles to handle sensitive distributed datasets efficiently.

    \subsubsection{CryptoQFL} In~\cite{chu2023cryptoqfl}, Chu et al. proposed CryptoQFL, which is a secure and efficient QFL framework that extends the baseline QFL by optimizing both computation and communication while maintaining strong security guarantees. In this framework, clients perform hybrid quantum-classical training on their local datasets, encoding the computed gradients into quantum states. Gradients are encrypted using a quantum one-time pad (QOTP), and key updates are performed locally, eliminating the need to transmit QOTP keys to the server, thereby reducing communication overhead and enhancing security. To address the large communication cost associated with high-precision gradient transfer, CryptoQFL employs ternary quantization, representing gradient values with three levels (-1, 0, 1) in a cyclic fashion suitable for quantum parameters. For efficient aggregation of these ternary gradients, a fast multi-bit quantum adder is designed that minimizes the use of costly non-Clifford gates, particularly Toffoli (CCX) gates, while eliminating garbage outputs and enabling serial operations on input qubits. These combined strategies significantly reduce computational latency, communication overhead, and resource consumption, allowing scalable, noise-resilient, and secure federated learning for quantum neural networks without compromising model convergence or accuracy.

    \section{Standardized Evaluation Metrics in Intrusion Detection}

    Evaluating FL approaches for network intrusion detection requires metrics that assess both model performance and system-level efficiency, while accounting for the unique aspects of QFL. Standardized evaluation facilitates a fair comparison between classical and quantum approaches and ensures practical relevance.

\subsection{Model Performance Metrics}
Applicable to both classical and quantum FL models:

    \textbf{Accuracy:} Overall proportion of correctly classified instances.
    \[
    \text{Accuracy} = \frac{TP + TN}{TP + TN + FP + FN}
    \]
    where \(TP\) = True Positives, \(TN\) = True Negatives, \(FP\) = False Positives, \(FN\) = False Negatives.

     \textbf{Precision, Recall, and F1-Score:} Essential for handling imbalanced intrusion detection datasets.
    \[
    \text{Precision} = \frac{TP}{TP + FP}, \quad
    \text{Recall} = \frac{TP}{TP + FN}, \quad\]
    \[F_1 = 2 \times \frac{\text{Precision} \times \text{Recall}}{\text{Precision} + \text{Recall}}
    \]

    \textbf{Area Under the ROC Curve (AUC-ROC):} Evaluates the trade-off between true positive rate (TPR) and false positive rate (FPR).
    \[
    \text{TPR} = \frac{TP}{TP + FN}, \quad
    \text{FPR} = \frac{FP}{FP + TN}
    \]
    The AUC is the area under the curve plotted with TPR against FPR.

    \textbf{Confusion Matrix Analysis:} Provides a detailed breakdown of classification results, showing \(TP\), \(TN\), \(FP\), and \(FN\) counts for each class. This helps identify which attack types are misclassified or under-detected.

\subsection{Federated Learning-Specific Metrics}
Metrics that capture distributed learning dynamics:

    \textbf{Communication Efficiency:} Number of communication rounds, data transmitted per round, and bandwidth utilization.
    
    \textbf{Convergence Rate:} Speed at which the global model stabilizes across heterogeneous clients.
    
    \textbf{Client Contribution and Fairness:} Ensures performance is balanced across clients with non-IID or uneven data distributions.
    
    \textbf{Robustness to Non-IID Data:} Measures model generalization when client datasets are heterogeneous.

\subsection{Resource and Operational Metrics}
Critical for deployment on real networks and edge devices:

    \textbf{Training and Inference Latency:} Time per local update and global aggregation.
    
    \textbf{Memory and Computation Requirements:} Particularly important for IoT or resource-constrained devices.

    \textbf{Energy Consumption:} Key for battery-powered or energy-limited nodes.

\subsection{Security and Privacy Metrics}
Essential for intrusion detection systems:

    \textbf{Data Privacy Guarantees:} Evaluates mechanisms like secure aggregation or differential privacy.
    
    \textbf{Adversarial Robustness:} Resistance to poisoning or inference attacks targeting FL.
    
    \textbf{Federated Model Fidelity (Quantum-Specific):} For QFL, measures how accurately quantum-encoded client states preserve class separability.
    
    \textbf{Quantum Kernel Separability:} Evaluates the ability of quantum feature maps to distinguish attack patterns in Hilbert space.

\subsection{Hybrid Considerations}
For systems combining classical and quantum FL:

\begin{itemize}
    \item Assess both classical performance metrics and quantum-specific metrics concurrently.

    \item Track additional resource usage from hybrid quantum-classical computations, including QPU access times and state preparation overhead.

    \item Monitor operational feasibility when integrating quantum-enhanced models with existing network infrastructure.
\end{itemize}

By using a combination of classical, FL-specific, and quantum-aware metrics, researchers and practitioners can perform holistic evaluations of federated intrusion detection systems, ensuring fair comparisons and guiding deployment decisions.

\section{Open Challenges in IDS}

The increase in digitization and online transactions has led to constant threats to the sensitive and personal data of customers. Also increase in IoT applications, in which sensors collect sensitive data has significantly heightened the risk of data breaches and compromises \cite{alazab2021federated}. There are various challenges for Federated Learning  (FL)  while implementing IDS \cite{agrawal2022federated,alazab2021federated}.

\textbf{Non-IID Data Distribution:} In FL, multiple clients (often referred to as edge nodes or devices) train models on their local data, which can vary significantly across different nodes. These edge nodes are typically distributed across diverse environments and collect data specific to their own context. As a result, the data recorded by these clients often exhibits non-Independent and Identically Distributed (non-IID) characteristics \cite{lu2024federated}. In other words, the data collected by each client may not follow the same distribution, leading to a heterogeneous data environment.

For instance, in the context of IDS, different edge devices might capture varying types of network traffic, depending on their role in the network, the region they are deployed in, and the specific network configurations they monitor. Some devices might focus on network traffic from IoT devices, while others might capture corporate network activity, resulting in diverse data distributions. This heterogeneity introduces several challenges in federated learning, particularly when trying to train a global model that generalizes well across all clients.

The non-IID nature of the data presents challenges in terms of model convergence, communication efficiency, and privacy \cite{li2022federated}. Since the data across devices differs, the gradients or model updates derived from these clients can vary significantly, causing difficulties in aggregating them effectively. In addition, certain clients might have more abundant data, while others may have limited data, further complicating the model's ability to learn generalized patterns.

Huang et al. \cite{huang2022federated} proposed a cluster-based method to deal with non-iid data in IDS. Unlike traditional FedAvg, which aggregates updates from individual nodes, this approach introduces cluster servers to group mutually trusted nodes into clusters. Each cluster operates as a single client in the FL framework. Within these clusters, local updates are aggregated before being passed to the central aggregation server, which then performs global model aggregation. This hierarchical structure significantly reduces the variability caused by non-IID data and enhances the scalability and security of the system. By applying this method, the performance of FL in IoT-based IDS is improved, as demonstrated using the IoT-23 dataset. In~\cite{ma2022state}, Ma et al. surveyed various techniques to deal with non-IID data in federated settings. The authors classified the techniques as data-based, model-based, datasets and data settings, algorithm-based and framework-based and provided future directions based on their survey.

The ability of quantum computing to encode classical data into exponentially high-dimensional Hilbert space can find complex patterns in non-linearly separable data. It can help find shared global patterns despite local heterogeneity. Quantum methods can approximate global gradients across clients with heterogeneous data more robustly~\cite{qi2023optimizing}.


\textbf{Heterogeneity:} Device heterogeneity is a significant challenge in FL, particularly in IoT environments where devices differ in computational power, memory capacity, and network connectivity. This disparity can lead to unequal contributions during the training process, as resource-constrained devices may struggle to complete local training tasks or communicate updates within the given time frame. Such inconsistencies can slow down the global model's convergence, reduce overall performance, and create bottlenecks in the system. Addressing device heterogeneity is essential to ensure fair participation of all devices and to maximize the benefits of FL in diverse IoT networks.

Asynchronous communication is a prominent approach, enabling devices to operate independently without synchronization, thereby mitigating the impact of stragglers caused by variability in computation and communication resources \cite{recht2011hogwild}. However, bounded-delay assumptions often used in asynchronous methods must be adapted to the unbounded delays typical in federated settings. Active sampling provides another solution by selectively choosing participating devices based on resource availability or statistical representativeness, ensuring efficient aggregation within time constraints \cite{nishio2019client}. Yet, extending such methods to handle real-time fluctuations in device performance remains an open challenge. Additionally, fault-tolerant mechanisms are critical, as device failures can bias the federated model if the dropped devices represent specific data distributions \cite{yin2018byzantine}. Strategies like ignoring failed devices, though simple, risk introducing bias, emphasizing the need for robust, adaptive techniques to balance fairness and efficiency across heterogeneous devices \cite{bonawitz2019towards}.

In~\cite{gurung2025communication}, Gurung et al. proposed a model-driven quantum federated learning algorithm (mdQFL), a novel framework to deal with device heterogeneity and non-IID data. Devices are grouped based on similarity between model parameters or performance metrics and a device selection method based on a selection criteria is proposed. The proposed framework shows the power of quantum computing in addressing device heterogeneity. Similar works on client grouping hardware-aware techniques can be found in~\cite{10623010, 10758814}.

\textbf{Threats to FL:}
\begin{itemize}
    \item \textbf{Inference Attacks:} These attacks in FL pose a significant threat to privacy, as adversaries can exploit shared gradients to extract sensitive information about participants' training data. Gradients, derived from private data, are computed using features and errors from deep learning model layers. Attackers can observe updates and infer details such as class representatives, membership information, or specific properties of the training data. In some cases, they can even reconstruct original training samples without prior knowledge of the dataset \cite{zhu2019deep}. These vulnerabilities arise because deep learning models often encode unintended features, allowing adversaries to infer private details by analyzing differences in consecutive model snapshots or aggregated updates. Such attacks highlight the critical need for robust privacy-preserving mechanisms in FL to protect participants' data.
    \item \textbf{Poisoning Attacks:} Poisoning attacks in FL aim to compromise the integrity of the model by altering its behavior in undesirable ways. These attacks can be random, seeking to reduce model accuracy, or targeted, aiming to misclassify specific inputs \cite{huang2011adversarial}. Poisoning can occur at two levels: data and model \cite{lyu2020threats}. Data poisoning involves manipulating the training data, either by subtly altering features without changing labels (clean-label attacks) \cite{shafahi2018poison} or by flipping labels to create mislabeled examples (dirty-label attacks) \cite{gu2017badnets}. For example, a malicious participant might flip the labels of all 1s to 7s, causing the model to misclassify 1s consistently. Model poisoning, on the other hand, involves directly tampering with local model updates or embedding hidden backdoors, making the model perform adversarial objectives when triggered by specific input patterns. Targeted model poisoning is particularly effective, as even a single malicious update can introduce backdoors \cite{li2024threats}. While data poisoning typically requires less sophistication, model poisoning demands higher technical expertise and resources. Both types of poisoning attacks highlight critical vulnerabilities in FL, especially when adversaries have the capability to manipulate training processes.
    \item \textbf{Sybil Attacks:} These attacks in FL involve an attacker creating multiple fake identities to amplify their influence on the global model. This makes FL vulnerable, especially since participants can freely join and leave during training. For example, research shows that just two Sybils can trick the model into misclassifying digits like “1” as “7" \cite{li2024threats}. Sybils can use similar or different datasets to craft malicious updates and often coordinate to overpower genuine clients. These attacks highlight a significant challenge in securing FL systems against manipulation.
\end{itemize}

To mitigate attacks in FL, several defense strategies are employed. Data sanitization filters malicious data before training by leveraging techniques like micro-model voting and global filters (e.g., FedDiv) \cite{cretu2008casting}. However, its reliance on direct data access raises privacy concerns, making it suitable only for trusted environments. Anomaly detection identifies irregularities in client behavior or data patterns through statistical or model-based methods, such as auto-encoders, spectral analysis, and visualization tools like VADAF \cite{chen2018autoencoder}. It helps exclude malicious clients and anomalous data but faces challenges with unsupervised scenarios and data access restrictions. Differential privacy (DP) ensures that individual data contributions remain indistinguishable, protecting sensitive information from leakage \cite{li2024threats}. Similarly, Secure Multi-Party Computation (SMPC) enables collaborative training without revealing data to other parties, offering robust privacy preservation \cite{li2024threats}. 

Threats to FL can be greatly minimized by the use of quantum techniques such as Quantum Key Distribution (QKD)~\cite{renner2008security} which ensure theoretically unbreakable encryption. Quantum Differential Privacy (QDP)~\cite{hirche2023quantum} techniques are being explored by researchers and can be future alternatives to the classical counterpart.

\textbf{Communication Overhead:} FL faces several challenges related to communication and computational efficiency \cite{shahid2021communication}. FL can face communication bottlenecks due to a huge number of participating devices, network bandwidth, and a limited amount of computational power as compared to central GPUs. While more devices can improve model training due to diverse data availability, managing many devices simultaneously may lead to communication bottlenecks and increased computational costs \cite{mothukuri2021survey}. FL reduces traditional ML costs, but reliable bandwidth is important. Unstable networks or mismatched upload/download speeds can delay model uploads, disrupting training. Edge devices, unlike powerful GPUs, have limitations in processing power, storage, and bandwidth. Training complex models on edge devices is significantly slower, even with advanced connectivity like 5G \cite{khraisat2024survey}.

In FL, local updating, client selection, decentralized training and peer-to-peer learning, model update and compression schemes reduction are critical techniques to improve communication efficiency \cite{shahid2021communication}. Briggs et al., in \cite{briggs2020federated}, the authors combine FL with Hierarchical Clustering (HC) to reduce communication rounds by grouping clients based on the similarity of their local updates to the global model. This approach, tested using the Manhattan distance metric, reduces the number of communication rounds.

Although classical data compression and asynchronous methods have mitigated some of the communication issues, emerging quantum advancements can bring a radical shift. Techniques such as quantum teleportation~\cite{pirandola2015advances}, and entanglement-assisted methods~\cite{hsieh2010entanglement} may enable ultra-efficient and secure transmission of model updates between clients and servers. Furthermore, quantum data compression~\cite{rozema2014quantum} and quantum gradient updates~\cite{kerenidis2020quantum,gilyen2019optimizing} are being explored that could reduce the communication footprint in QFL.

While quantum computing seems to provide solutions to most of the challenges faced by classical deep learning methods in federated settings, it is to be noted that quantum computation has its own challenges and is still in the developing phase. Integrating quantum methods into FL introduces a new layer of complexities. In the real world, quantum computation has a lot of constraints such as hardware limitations, including limited qubits, short coherence, and susceptibility to noise. Moreover, in FL settings, novel hybrid architectures need to be developed for seamless interaction between classical edge devices and quantum resources (such as communication protocols, and intelligent schedulers). 

Beyond these technical concerns, practical implementation challenges arise when moving from research to deployment:

\textbf{Deployment Costs:} Implementing advanced IDS, especially those that use federated learning and quantum-enhanced methods, can produce significant costs. These costs include acquisition of various hardware(such as QPUs, GPUs, or TPUs), network infrastructure and licenses for commercial software frameworks. Smaller organizations and resource-constrained environments may find these requirements prohibitive.

\textbf{Maintenance and Operational Complexity:} Continuous updates are required to detect evolving attacks. In addition, maintenance overhead is introduced in a federated setting that includes handling client failures, ensuring efficient communication between the clients and the server, and ensuring synchronization between devices within a network. Quantum models may require recalibration of encoding circuits or tuning of hybrid classical-quantum training loops, adding further operational complexity.

\textbf{Integration with Existing Security Infrastructure:} Deploying new IDS models alongside legacy systems can be challenging. Compatibility with existing network monitoring tools, SIEM platforms, firewalls, and logging frameworks must be ensured. Data formats, protocols, and storage mechanisms may need adaptation, and real-time alerting pipelines must be carefully designed to prevent bottlenecks or false alarms.

\textbf{Scalability and Resource Management:} High-volume networks demand scalable solutions. FL models must manage communication overhead, model aggregation latency, and client heterogeneity. Quantum-enhanced solutions are currently limited by qubit count and coherence time, constraining their applicability to large-scale deployments.

\textbf{Regulatory and Privacy Considerations:} Federated learning mitigates some privacy risks, but compliance with data protection regulations (e.g., GDPR) still requires careful design. Quantum approaches introduce additional considerations related to data encoding and secure state preparation, particularly when sensitive network traffic is involved.

\section{Regulatory Considerations}

\subsection{Regulatory Compliance}

Deploying FL-based IDS in real-world networks introduces important regulatory, ethical, and societal considerations that go beyond technical performance. Addressing these concerns is critical to ensure compliance, maintain public trust, and maximize the societal benefits of such systems.

Organizations deploying FL-based NIDS must comply with local and international data protection and cybersecurity regulations. Key considerations include:

\textbf{Data Privacy Regulations:} Laws such as the General Data Protection Regulation (GDPR)~\cite{GDPR} in Europe, the California Consumer Privacy Act (CCPA)~\cite{CCPA}, and national cybersecurity frameworks impose strict requirements on collection, storage, and sharing of personal and sensitive data. FL partially mitigates privacy risks by keeping raw data on local devices, but mechanisms like differential privacy and secure aggregation are necessary to comply with regulatory standards.

\textbf{Industry-Specific Standards:} Critical infrastructure operators, financial institutions, and healthcare organizations may need to adhere to sector-specific regulations (e.g., NIST Cybersecurity Framework, ISO/IEC 27001) when integrating FL into intrusion detection pipelines~\cite{NIST, ISO}.

\textbf{Audit and Accountability:} Regulators increasingly require transparent reporting of algorithmic decisions. FL models must provide traceable logs of training, aggregation, and deployment steps to ensure accountability.

\subsection{Ethical Considerations}

Ethical deployment of FL-based NIDS encompasses fairness, transparency, and user consent:

\textbf{Fairness Across Clients:} FL systems often operate across heterogeneous networks with varying data distributions. Ethical deployment requires that model performance is balanced across all participants, avoiding bias toward certain clients or network segments.

\textbf{Transparency and Explainability:} Security teams must understand why a model flags specific traffic as malicious. Explainable AI (XAI)~\cite{nwakanma2023explainable} methods are critical to maintain trust in automated detection systems.

\textbf{Consent and Awareness:} Users and organizations contributing data to FL should be informed about what data is used, how it is processed, and the potential risks and benefits.

\subsection{Societal Impacts}

FL-based NIDS can have broad societal implications:

\textbf{Enhanced Cybersecurity:} Widespread deployment can reduce successful attacks and mitigate threats across industries, protecting critical infrastructure and sensitive information.

\textbf{Privacy Preservation:} By avoiding centralization of sensitive network data, FL reduces exposure risks, potentially increasing public trust in digital services.

\textbf{Inequity Risks:} Organizations with fewer resources may be unable to participate in FL networks, potentially widening the cybersecurity gap.

\textbf{Dependence on AI Systems:} Over-reliance on automated intrusion detection without human oversight could result in unforeseen consequences, such as misclassifications or systemic vulnerabilities.

\section{Industrial Adoption Roadmap}

\begin{figure*}[ht]
    \centering
    \includegraphics[width=1.00\linewidth]{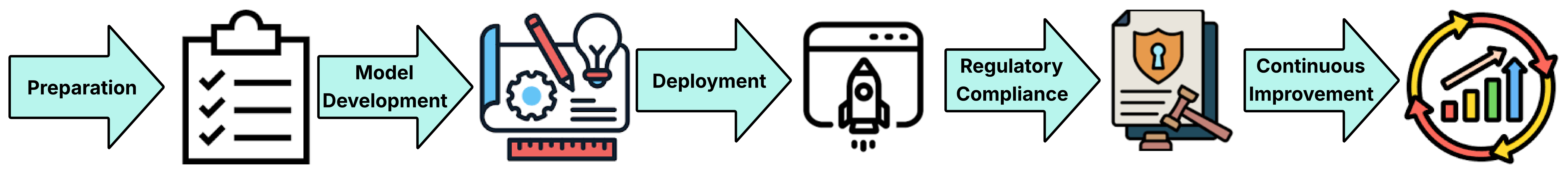}
    \caption{Roadmap illustrating the pathway for industrial adoption of FL-based IDS, highlighting the key phases from initial planning to operational deployment and ongoing optimization.}
    \label{fig:roadmap}
\end{figure*}

The industrial adoption of FL-based IDS can be achieved in the following five phases (Fig.~\ref{fig:roadmap}):

\begin{enumerate}
    \item Preparation: It is one of the key steps to develop secure and effective FL-based IDS. It involves ensuring that distributed datasets exist across devices or sites, handling preprocessing, normalization, and feature selection, and establishing the required infrastructure by deciding between on-premises, edge, or cloud-based deployment while also ensuring secure networking and SDN compatibility.
    \item Model Development: The second phase focuses on selecting appropriate federated learning algorithms such as FedAvg, FedProx, or FedGAN, deciding between centralized or decentralized aggregation strategies, designing intrusion detection models (e.g., GNNs, autoencoders, CNNs), and defining standardized evaluation metrics such as accuracy, F1-score, and detection latency.
    \item Deployment \& Communication: The third phase emphasizes the execution of federated training at client sites, secure communication through techniques such as homomorphic encryption and differential privacy, and the implementation of appropriate aggregation strategies, either server-side or hierarchical/peer-to-peer.
    \item Regulatory \& Ethical Alignment: The fourth phase ensures compliance with regulations such as GDPR, HIPAA, or CCPA, enforces data minimization and privacy guarantees, and establishes trust and governance through transparency in aggregation and accountability frameworks.
    \item Standardized Evaluation \& Continuous Improvement: Finally, the fifth phase highlights the use of standardized benchmarks such as UNSW-NB15~\cite{7348942} and Bot-IoT~\cite{koroniotis2019towards}, defines cross-industry IDS benchmarks, and stresses continuous model re-training to handle evolving cyber threats while enabling real-time anomaly detection.
\end{enumerate}

\section{Future Directions}

With the rapid advancements in AI, FL has huge potential for revolutionizing privacy-preserving and distributed computing for applications like intrusion detection and anomaly detection. We present a few potential research directions for FL in IDS.

To fully realize the potential of FL, several key challenges need attention. Enhancing privacy-preserving techniques is of utmost importance. Approaches like advanced differential privacy mechanisms, zero-knowledge proofs, and blockchain integration~\cite{pokhrel2020federated} can protect sensitive data while maintaining system integrity. For instance, blockchain can ensure the immutability and reliability of model updates \cite{8843900, xu2022fair}, while quantum-resistant cryptography~\cite{renner2008security} can prepare FL for future challenges posed by quantum computing \cite{li2024enhancing}.

Optimizing computational and communication efficiency is crucial, especially for large-scale IDS deployments. Novel optimization algorithms that accelerate convergence, reduce the computational burden, and minimize communication overhead are vital. Techniques like selective data sharing, model compression \cite{shah2021model}, and secure aggregation can address these challenges without compromising security. Current quantum hardware is limited by noise, qubit count, and coherence times. Developing QFL algorithms and error-mitigating techniques for NISQ devices can be a great innovation in the real world IDS deployment. Designing effective quantum feature encoding techniques particularly tailored to capture complex network traffic patterns can enhance IDS.

While QFL provides privacy benefits, quantum-specific threats and vulnerabilities (such as quantum adversarial attacks)~\cite{liao2021robust} require new quantum-safe privacy preservation methods to be developed. 

Since the network topology is dynamic and constantly changing in the real world, representing network data in graphs can be very effective as it captures inter-device communication patterns and facilitates the application of graph learning methods. Graph learning based on QML has received limited attention specifically in IDS, providing a valuable opportunity for future research.

To be effective, FL should accommodate the variability and dynamics of real-world settings. IDS systems often face concept drift due to evolving attack patterns and data distributions. Online learning, continual learning, and transfer learning offer promising solutions by enabling models to adapt in real time without losing previously learned knowledge. Integrating real-time threat intelligence and dynamic feedback mechanisms with QML can further enhance IDS responsiveness.

By addressing these challenges, FL can become a scalable, robust, and secure solution for modern IDS, ensuring effective protection against emerging cyber threats in diverse and evolving environments.

\section{Conclusion}

This survey provides a thorough examination of the integration of Federated Deep Learning (FDL) and Quantum Federated Learning (QFL) in NIDS. As the cyber threats evolve, there is a need for robust and scalable solutions for network security. By exploring a wide range of techniques, trends, and methodologies, this paper highlights the significant advancements in leveraging federated learning to enhance the privacy, scalability, and performance of intrusion detection systems. This survey provides a comprehensive overview of the field by exploring essential topics, including privacy concerns, key challenges, potential threats, aggregation strategies, communication methods, and practical deployment scenarios. By addressing these aspects, it offers a well-rounded perspective on the latest developments and advancements in the domain.

This survey is up-to-date and covers all the latest trends and technologies that are relevant to NIDS. Unlike existing papers, which may lack certain critical information, this survey provides a comprehensive overview of the topic, ensuring that no key aspects are overlooked. Researchers can rely on this work as a solid foundation to further their studies, explore novel ideas, and build upon the current state of knowledge in the field.

\bibliographystyle{IEEEtran}
\bibliography{ref}

\end{document}